\documentclass[%
 reprint,
superscriptaddress,
nofootinbib,
 amsmath,amssymb,
 aps,
pra,
]{revtex4-2}

\usepackage[margin=2cm]{geometry}
\usepackage{graphicx} % Required for inserting images
\usepackage{placeins}
\usepackage[utf8]{inputenc}
\usepackage[utf8]{inputenc}      % input font encoding
\usepackage[colorlinks=true,citecolor=blue,urlcolor=blue,linktocpage=true,linkcolor=blue]{hyperref}
\usepackage{amsmath,amssymb,mathtools}
\usepackage[T1]{fontenc}          % output font encoding
\usepackage{booktabs,tabularx}
\usepackage{xspace}
\usepackage[usenames]{xcolor}%\definecolor{fscolor}{RGB}{44,118,255}
\usepackage{geometry}
\usepackage{xparse}
\usepackage{bbm}
\usepackage{tabularx}
\usepackage{soul}
\usepackage[capitalize]{cleveref}
\usepackage{slashed}
\usepackage{braket}
\usepackage{leftindex}
\usepackage{pifont}
\usepackage[shortlabels]{enumitem}
\usepackage{subfigure}
\usepackage[super]{nth}
\usepackage{bm}
\usepackage{lineno}

\usepackage{siunitx}
\usepackage{makecell}

\begin{document}
\title{Resonant photon scattering in the presence of external fields\\ and its applications for the Gamma Factory}

\author{Jan Richter}
\email{jan.richter@ptb.de}
\affiliation{Physikalisch-Technische Bundesanstalt, Bundesallee 100, 38116 Braunschweig, Germany}
\affiliation{Institut für Theoretische Physik, Leibniz Universität Hannover, Appelstraße 2, 30167 Hannover, Germany}

\author{Mieczyslaw Witold Krasny}
\affiliation{LPNHE, Sorbonne Université, Université de Paris, CNRS/IN2P3, Tour 33, RdC, 4, pl. Jussieu, 75005 Paris, France}
\affiliation{CERN, BE-ABP, 1211 Geneva 23, Switzerland}
		
\author{Jan Gilles}
\affiliation{Technische Universität München, D-80333 München, Germany}
\affiliation{Ludwig-Maximilians-Universität München, D-80539 München, Germany}

\author{Andrey Surzhykov}
\affiliation{Physikalisch-Technische Bundesanstalt, Bundesallee 100, 38116 Braunschweig, Germany}
\affiliation{Institut f\"ur Mathematische Physik, Technische Universität Braunschweig, Mendelssohnstrasse 3, D-38106 Braunschweig, Germany}

\begin{abstract}
    We present a theoretical study of resonant elastic photon scattering by highly charged ions in the presence of external electric and magnetic fields.
    Special emphasis is placed on head-on collisions of relativistic ion beams and counter-propagating laser photons, as can be observed in ion storage ring experiments.
    If the collision zone for such a scenario is exposed to a moderate field of a dipole magnet, a fast moving ion will be subjected to strong electric and magnetic fields in its rest frame due to the Lorentz transformation.
    To investigate the effect of these fields we performed detailed calculations for the elastic photon scattering by He-like Ca ions moving with the Lorentz factor $\gamma_L = 2395$.
    Such a scenario 
    %MWK attracts 
    is of 
    %MWK
    particular interest for the Gamma Factory project which is
    %MWK discussed 
    supported by  
    %MWK 
    the CERN Physics Beyond Colliders framework.
    Based on the results of our calculations, we argue that even moderate magnetic fields significantly modify the rate of detected photons as well as their emission pattern and polarization.
    This sensitivity to external fields opens interesting avenues for diverse applications for the Gamma Factory project, such as resonance condition tuning, beam cooling, and polarization control.    
    % This theoretical study explores resonant elastic photon scattering in the presence of external electric and magnetic fields, motivated by potential applications in storage ring experiments, such as the Gamma Factory project at CERN. In this framework, resonant scattering involves head-on collisions of relativistic ion beams and counter-propagating laser photons, leading to a strong enhancement of the external field strength due to the Lorentz transformation between ion rest and laboratory frame. Calculations for He-like Ca ions reveal a notable impact of the external fields on the scattering rate as well as the angular distribution and polarization of emitted photons. This opens interesting avenues for diverse applications for the Gamma Factory project, such as as resonance condition tuning, beam cooling, and polarization control.
\end{abstract}

\maketitle

\section{Introduction}\label{Section:Introduction}

Elastic photon scattering by atoms and ions is one of the most fundamental processes of light-matter interaction and has been the subject of numerous experimental and theoretical studies~\cite{kane1986,Middents2023,Strnat2021,Strnat2022,Serbo22,Volotka22,Samoilenko2020}. 
Special attention in these studies is given to the scattering of photons, whose energy is close to the transition energy between the ground and some excited atomic state.
This so-called resonant photon scattering is of interest for various applications such as, for example, measurements of oscillator strengths and lifetimes of highly charged ions~\cite{Bernitt2012,Togawa2024,richter2024}, investigations of spacial modulations and electronic structure of complex materials~\cite{Fink2013}, and the production of high-intensity $\gamma$-rays.
%MWK see small modifications below
The latter case is expected to be realized by the Gamma Factory project at CERN~\cite{Krasny2015}, which is supported by the Physics Beyond Colliders framework~\cite{Jaeckel2018}.
The key principle of the Gamma Factory is rooted in the resonant scattering of relativistic ion beams and counter-propagating laser photons involving a strong enhancement of the photon frequency caused by the Doppler boost.
The use of Doppler boosted laser photons is well established in atomic physics experiments and was successfully employed already in the 1970s~\cite{Bryant77,Bryant78}.
For ultra-relativistic energies of the projectile ions, the resonant scattering can produce small-divergence, high-energy photon beams with tremendous potential for a wide range of experiments~\cite{Apyan:2022ysh, Budker:2021fts,Balkin:2021jdr, Nichita:2021iwa,  Budker2020, Zimmermann:2022svn, Chakraborti:2022ubp, Karbstein:2021otv,Chakraborti:2021hfm, Wojtsekhowski:2021xlh, Flambaum:2020bqi,Baolong:2024ata}.\\
A deep understanding of the resonant photon scattering is needed for planning and guidance of future Gamma Factory experiments.
A number of theoretical studies have been reported recently, therefore, that analyzed the angular distribution and polarization of scattered photons~\cite{Samoilenko2020,Serbo22,Volotka22}.
Those studies have been performed within the fully relativistic framework and by taking into account the electron-photon interaction beyond the dipole approximation, but in the absence of external (static) electric and magnetic fields.
Thus, the previous theoretical analysis can be used to describe a typical Gamma Factory setup where the light-ion interaction takes place in the field-free zone of the storage ring. 
In this contribution, we present an extension of previous theoretical studies towards resonant photon scattering in the presence of external fields.
These fields, as seen in the rest frame of the ion, can be introduced by a dipole magnet installed in the collision zone.
Due to the Lorentz boost, already a weak laboratory magnetic field can have a strong impact on the scattering process.
Such Lorentz transformations of external fields have been successfully applied in the past for beam energies much smaller than those available at the LHC~\cite{Bryant78,Harris90,Harris93}.
In the present study we will deal with projectile velocities approaching the speed of light, leading to strong Lorentz-boosted fields that open up new possibilities for the Gamma Factory project.

To analyze the effects of external fields on the scattering process, one needs to specify the geometry of the ``ion \textit{plus} light'' system exposed to the external electric and magnetic fields in both the laboratory and the ion frame.
The geometry is discussed in Section~\ref{Section:Geometry}, together with basic parameters used in our study.
After the geometry is established, we turn to the analysis of the influence of external fields both on structure and dynamics of ions.
In Section~\ref{Section:EnergyShifts}, for example, we briefly recall Zeeman and Stark shifts that modify the electronic levels of an ion.
The scattering amplitude, which is the main ``building block'' in the theory of resonant scattering, is obtained later in Section~\ref{Section:ResonantScattering} by taking the effects of external fields into account.
In Section~\ref{Section:He-like Ions} we apply the derived theory to the description of the $1\text{s}^2$~$^1\text{S}_0$ $ \to$ $1\text{s}2\text{p}$~$^1\text{P}_1 $ $\to$ $1\text{s}^2$~$^1\text{S}_0$ photon scattering by He-like Ca ions.
Here, we pay special attention both to the reaction rate and to the angular distribution and polarization of scattered photons.
We argue that even moderate fields of a dipole magnet (in the laboratory frame) can significantly modify these properties.
Besides the fundamental interest, the high sensitivity of resonant photon scattering to external fields offers many interesting applications for the Gamma Factory project.
These applications, such as resonance condition tuning, beam cooling, and polarization control, are discussed in Section~\ref{Section:Applications}.
Finally, the most important aspects of the present study will be summarized in Section~\ref{Conclusion}.

\section{Geometry and basic parameters}\label{Section:Geometry}

\begin{figure} 
\includegraphics[width=0.45\textwidth]{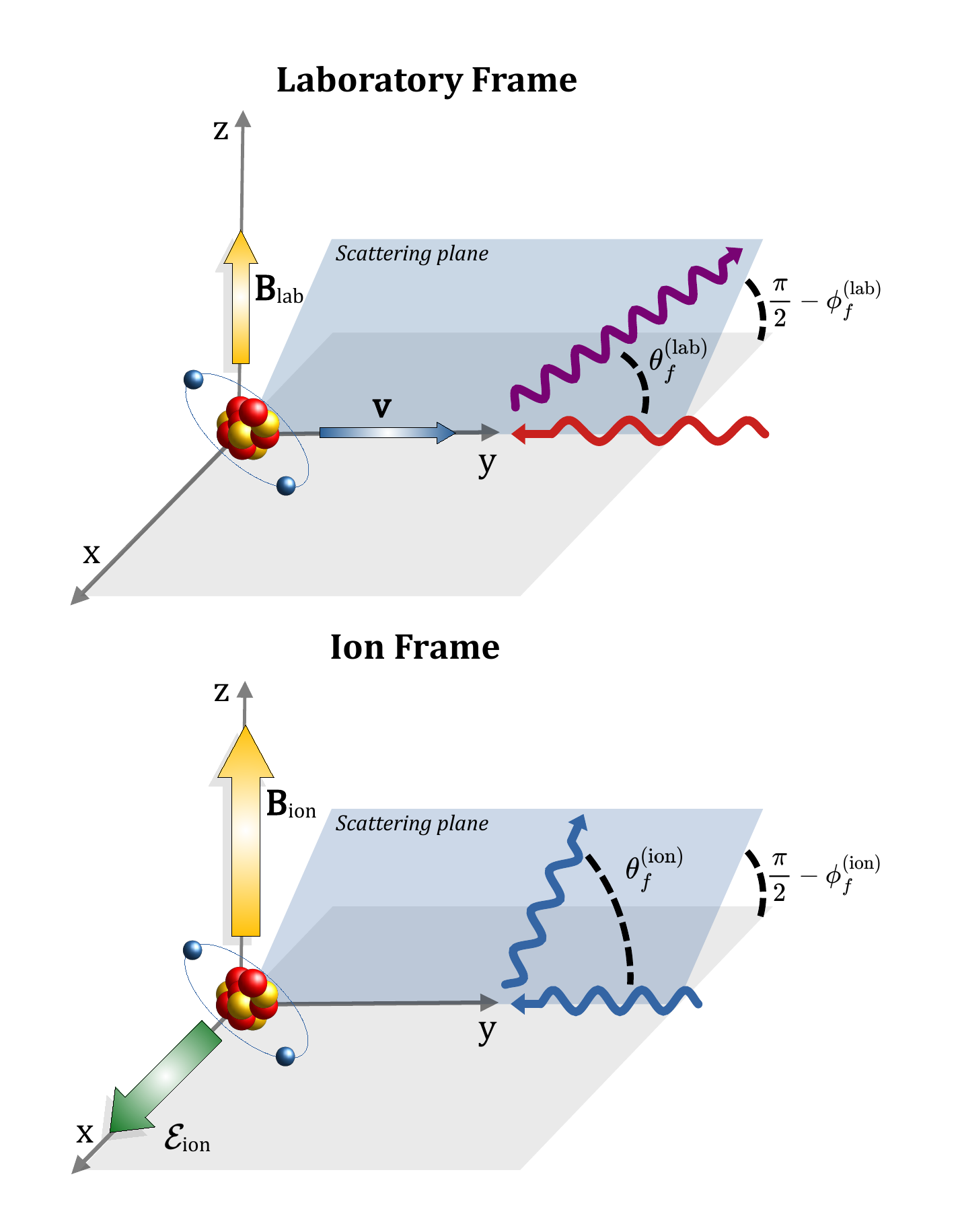}
\caption{\label{fig:Field-Geometry} Elastic photon scattering by highly charged ions in the laboratory (upper panel) and ion rest frame (lower panel). The direction of the ion beam propagation in the laboratory frame defines the y-axis which together with the magnetic field spans the y-z-plane. While the incident photons counter propagate to the ion beam, the emission direction of the outgoing photons is characterized by the angles $\theta_f$ and $\phi_f$. Due to Lorentz transformation an electric field emerges in x-direction in the ion frame.}
\end{figure}

Before we delve into the theory of resonant scattering we have to discuss the geometry of this process. For the case of a head-on collision between photons and a relativistic ion beam, which is a typical scenario for storage ring experiments, the geometry is shown in Fig.~\ref{fig:Field-Geometry}. 
Here, the ions move along the y-axis while the photons counter propagate in negative y-direction. The direction of the emitted photon is characterized by the polar scattering angle $\theta_{f}$ and the azimuthal angle $\phi_f$ which are defined with respect to the y-axis and the y-z-plane, respectively.\\ 
Taking into account the relativistic motion of the ions, the scattering angle depends on the choice of a particular reference frame. In this study, two reference frames are considered: the laboratory frame and the ion rest frame. All calculations are performed in the latter, and the results are then Lorentz transformed into the laboratory frame, where all measurements take place.
This transformation can be easily carried out by using the basic relations for the scattering solid angles:
\begin{equation}
 \cos \theta_f^{(\mathrm{ion})} = \frac{\cos \theta_f^{(\mathrm{lab})}-\beta}{1-\beta \cos \theta_f^{(\mathrm{lab})} }, 
 \label{eq.: lorentz trafo angle}
 \end{equation}
 and
 \begin{equation}
 \frac{\mathrm{d}\Omega^{(\mathrm{ion})}}{\mathrm{d}\Omega^{(\mathrm{lab})}} = \frac{1-\beta^2}{\left(1-\beta \cos \theta_f^{(\mathrm{lab})} \right)^2},
 \label{eq.: lorentz trafo solid angle}
 \end{equation}
with $\beta = v/c$ and the ion beam velocity $v$~\cite{eichler1995relativistic}.\\

In the present study, special attention will be paid to the effect of external electromagnetic fields on the scattering process. 
Below, we will assume that the ion beam in the laboratory frame is exposed to a static external magnetic field orthogonal to the ion velocity (see upper panel of Fig.~\ref{fig:Field-Geometry}). This corresponds to a scenario in which the ion beam propagates through the straight collision zone of the storage ring and is exposed to the field of a dipole magnet. 
For the entirety of this paper, the direction of this magnetic field defines the z-(quantization)-axis.
Due to the relativistic motion of the ions, the laboratory magnetic field $B_\mathrm{lab}$ gives rise to both magnetic and electric fields in the ion frame:
% We consider a realistic storage ring scenario in which the ion beam moves perpendicular to an external magnetic field generated by a dipole magnet in the collision zone (see Section~\ref{Section:Applications}).
% In the laboratory frame, this magnetic field defines the Z-(quantization)-axis.\\
% Due to the relativistic motion of the ions, the dipole field of the magnet gets Lorentz-transformed in the ion frame as
\begin{subequations}
\label{Eq.: Lorentz transformation fields}
\begin{align}
\bm{B}_{\mathrm{ion}} & =\gamma_L \bm{B}_{\mathrm{lab}}, \label{Eq.: Lorentz transformation B}\\
\bm{\mathcal{E}}_{\mathrm{ion}} & =\gamma_L \left(\bm{v} \times \bm{B}_{\mathrm{lab}}\right).\label{Eq.: Lorentz transformation E}
\end{align}
\end{subequations}
 As seen from Eq.~\eqref{Eq.: Lorentz transformation B}, the direction of the magnetic field does not change under transformation while its strength is enhanced by the Lorentz factor $\gamma_L=1/\sqrt{1-\beta^2}$. Moreover, in the ion frame an electric field emerges in x-direction with a field strength also proportional to $\gamma_L$. This configuration of the fields is shown in the lower panel of Fig.~\ref{fig:Field-Geometry}.\\
 For further discussions, it is helpful to gain an understanding of the magnitude of the field strengths within the ion frame.
 In Tab.~\ref{tab:Field Strength} we display $B_\mathrm{ion}$ and $\mathcal{E}_\mathrm{ion}$ for the case of $B_{\mathrm{lab}}=1$ T and different Lorentz factors achievable in the SPS and LHC~\cite{Budker2020}. 
 It is informative to compare the resulting values of $\mathcal{E}_\mathrm{ion}$ with the electric field strengths experienced by an electron in the ground state of a neutral hydrogen atom ($\mathcal{E}\approx 5\times 10^{11}$ $\text{ Vm}^{-1}$) and a hydrogen-like calcium ion ($\mathcal{E}\approx 4\times 10^{15}$ $\text{ Vm}^{-1}$).
As seen from this comparison the Lorentz-boosted external electric field is more than thousend times weaker compared to the Coulomb field in the Calcium ion, which will be employed as a testbed in our study.
 In order to question the strength of the external magnetic field it is natural to compare the fine-structure splitting and Zeeman shift of ionic levels.
 For the H-like Ca ion these are $\Delta E_{2p_{1/2} - 2p_{3/2}} \approx 7$ eV and $\Delta E_Z \sim \mu_B B_\mathrm{ion}\approx 0.2$\,eV for $B_\mathrm{lab}=1$\,T and $\gamma_L = 3000$, where $\mu_B$ denotes the Bohr magneton.
 We can conclude, therefore, that effects of external electromagnetic fields on medium- and high-Z highly charged ions can be treated perturbatively.
 
 % The resulting values of $B_\mathrm{ion}$ and $\mathcal{E}_\mathrm{ion}$ are compared to the electric field strength experienced by an electron in the ground state of a neutral hydrogen atom and a hydrogen-like calcium ion.
 To describe a realistic scenario, we will assume moreover that the ions exhibit a finite spread of longitudinal velocity and hence momentum. As usual in accelerator physics, this distribution is characterized by the parameter $\Delta p/p$, where $\Delta p$ is the uncertainty of the momentum. In the analysis below, we will consider two scenarios, which correspond to an ion beam injected without beam cooling and one exposed to additional laser cooling~\cite{Eidam:2017csp, Krasny:2020wgx,Kruyt:2024sty} (see Sec.~\ref{Sec: Beam cooling}). These two scenarios will be referred to as uncooled and cooled beams and are characterized by the values $\Delta p/p\approx 2\times 10^{-4}$ and $\Delta p/p\approx 1\times 10^{-5}$, respectively~\cite{Krasny:2020wgx}.

{\renewcommand{\arraystretch}{1.4}
\begin{table}[]
    \centering
    \renewcommand{\arraystretch}{1.5}
    \begin{tabular}{lll}
    \hline \hline
    &$B_\mathrm{ion}$ (T)\hspace{0.7cm} &$\mathcal{E}_\mathrm{ion}$ ($\text{ Vm}^{-1}$)\\ \hline
        $\gamma_L = 15$ & $15$ &$4 \times 10^{9}$ \\
        $\gamma_L=100$ & $100$ &$3 \times 10^{10}$ \\
        $\gamma_L=3000$\hspace{0.7cm} & $3000$ &$9 \times 10^{11}$ \\
        \hline \hline
    \end{tabular}
    \caption{Electric and magnetic field strengths in the ion frame  for a laboratory magnetic field strength of $B_{\mathrm{lab}} = 1$~T.}
    \label{tab:Field Strength}
\end{table}
}
\section{Energy Shifts induced by External Fields}\label{Section:EnergyShifts}
In order to analyze the effect of external electromagnetic fields on resonant photon scattering, one has to understand how such fields affect the electronic structure of an ion.\\
For an ion exposed to a strong magnetic field, the linear and quadratic Zeeman shifts have to be taken into account. If moreover an electric field is applied, the ionic energy levels are altered by the quadratic Stark shift. 
Below, we will briefly present the basic equations of the Zeeman and Stark shifts.

% this section gives a brief overview of the considered effects on the atomic structure, which are both  the linear and quadratic Zeemanshift, caused by the magnetic field, and the quadratic Stark shift, caused by the electric field.
%
%
%
%
%
\subsection{Zeeman Shift}\label{Section:Zeeman}
The interaction of a bound electron, carrying a magnetic moment $\bm\mu$, with an external magnetic field $\bm{B}$ is described by the operator
\begin{equation}
    \hat{H}_Z=-\bm\mu \cdot \bm{B}.
    \label{Eq.: Zeeman interaction Hamiltonian}
\end{equation}
In first order perturbation theory, this interaction leads to a shift of atomic energy levels which is linear in the magnetic field strength $B=\left|\bm{B}\right|$: 
\begin{equation}
    \Delta E_Z^{(1)} = \mu_B g B M_J.
    \label{Eq.: Zeeman first order}
\end{equation}
Here, $M_J$ is the projection of the atomic state angular momentum in z-direction and $g$ is the Landé $g$-factor.
Eq.~\eqref{Eq.: Zeeman first order} leads to the well known splitting of a state with total angular momentum $J$ into $2J+1$ magnetic sublevels.

Treating the Hamiltonian from Eq. \eqref{Eq.: Zeeman interaction Hamiltonian} in second order perturbation theory, one obtains the additional quadratic Zeeman shift
\begin{equation}
    \Delta E_{Z}^{(2)} = C_2 B^2,
\end{equation}
which is proportional to $B^2$.
For more details about the calculation of the quadratic Zeeman shift coefficient $C_2$, see Ref.~\cite{Gilles2024,Lu_2022}.

 %The QZSC for a state $|\Gamma J \rangle$ can be expressed in terms of reduced matrix elements of the magnetic dipole operator $\mu^{(1)}$

% \begin{equation}
%     C_2 = \frac{1}{2J + 1} \sum_{\Gamma', J'} \frac{|\langle J M; 1 0 | J' M\rangle|^2}{E(\Gamma J) - E(\Gamma' J')} |\langle \Gamma' J'||\mu^{(1)} + \Delta \mu^{(1)} ||\Gamma J\rangle|^2
%     \label{Eq.: Quadratic Zeeman shift coefficient}
% \end{equation}

%where $J$ is the total angular momentum, $\Gamma$ denotes the set of additional quantum numbers needed to uniquely specify the state, $M$ is the projection and $E(\Gamma J)$ the energy of the given state.
%The summation in (\ref{Eq.: Quadratic Zeeman shift coefficient}) in principle goes over the entire atomic spectrum. However, excellent approximations can be achieved by only summing over energetically nearby intermediate states $|\Gamma' J' \rangle$.
%
%
%
%
%
\subsection{Stark Shift}\label{Section:Stark}
Similar to the Zeeman effect, the interaction of a bound electron with an external electric field $\bm{\mathcal{E}}$, well known as the Stark effect, is described by the operator
% ~\cite{Angel1968,Maltsev2023,Yerokhin2016}:
\begin{equation}
   \hat{H}_S = e\bm{r}\cdot\bm{\mathcal{E}}.
\end{equation}
 As shown in Refs.~\cite{Angel1968,Yerokhin2016}, the quadratic Stark shift for a state with total angular momentum $J$ and projection $M_J$ along the quantization z-axis is given by  
\begin{align}
    \Delta E_S =& -\frac{1}{2} \alpha_0 \mathcal{E}^2 \nonumber \\
    &- \frac{1}{4} \alpha_2 \frac{3 M_J^2-J(J+1)}{J(2 J -1)}(3\mathcal{E}_z^2 - \mathcal{E}^2 ).
    \label{Eq.: Stark Shift}
\end{align}
Here, $\alpha_0$ and $\alpha_2$ are the scalar and tensor polarizabilities which depend on the atomic state but not on its angular momentum projection $M_J$. Hence, only the second term in Eq.~\eqref{Eq.: Stark Shift} introduces a splitting of sublevels with different values of $M_J^2$. For a more thorough analysis of the scalar and tensor polarizabilities see Refs.~\cite{Angel1968,Yerokhin2016}.

\section{Resonant Scattering}\label{Section:ResonantScattering}
This section aims to revisit the basic theory of resonant elastic photon scattering, whose detailed analysis is given in Refs.~\cite{Serbo22,Volotka22,Samoilenko2020}. First, we will discuss the scattering amplitude and pay special attention to the effects of the Zeeman and Stark shift. By making use of this amplitude and the well known density matrix approach we evaluate the scattering cross section and polarization properties of the emitted photons.

\subsection{Scattering Amplitude}

% The elastic scattering of photons on bound atomic electrons is represented by the second order matrix element~\cite{kane1986}
% \begin{align}
% \label{Eq.: Scattering matrix element}
%     &\mathcal{M}_{M_f,M_i}\notag\\
%     &=\alpha \sum_{\nu}\left[\frac{\braket{f|\hat{\mathcal{R}}^\dagger\left(\bm{k}_f,\bm{\epsilon}_f\right)|\nu}\braket{\nu|\hat{\mathcal{R}}\left(\bm{k}_i,\bm{\epsilon}_i\right)|i}}{E_i-E_\nu+\omega_i}\right.\notag\\
%     &\left.+\frac{\braket{f|\hat{\mathcal{R}}\left(\bm{k}_i,\bm{\epsilon}_i\right)|\nu}\braket{\nu|\hat{\mathcal{R}}^\dagger\left(\bm{k}_f,\bm{\epsilon}_f\right)|i}}{E_i-E_\nu-\omega_i}\right],
% \end{align}
% with the initial state $\ket{i}$ and the final state $\ket{f}$. Moreover, Eq.~\eqref{Eq.: Scattering matrix element} contains a sum over the intermediate state $\ket{\nu}$ which runs over the complete atomic spectrum including the positive and negative continuum. The operators $\hat{\mathcal{R}}$ and $\hat{\mathcal{R}}^\dagger$ describe the absorption and emission of a photon. The matrix elements of these operators are usually evaluated by using a multipole expansion and the Wigner-Eckart theorem (see e.g.~\cite{}). \\
\begin{figure}
    \centering
    \includegraphics[width=0.45\textwidth]{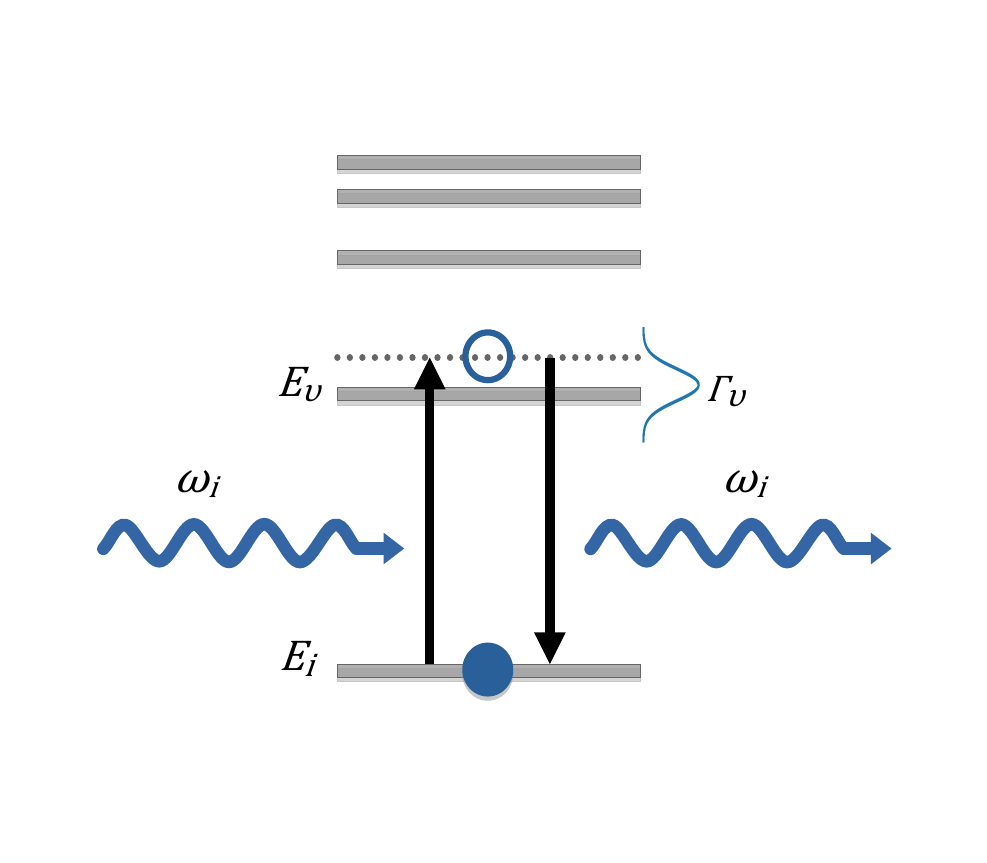}
    \caption{Schematic illustration of the resonant scattering process. The electron is excited from the ground state to the intermediate state with the natural width $\Gamma_\nu$ by absorbing the incoming photon with an energy of $\omega_i \approx E_\nu - E_i$. Afterwards a second photon is emitted with $\omega_i=\omega_f$.}
    \label{fig:Resonant Scatteirng schematic}
\end{figure}
For low intensities, the coupling between light and an ion is usually treated within perturbation theory. For the process of photon scattering, this leads to the well known second order scattering amplitude, thoroughly discussed in the literature~\cite{kane1986,Strnat2021,Strnat2022,Manakov_2000,Serbo22}.
In the resonant case, when the photon energy $\omega_i$ approaches the excitation energy of some intermediate ionic state, see Fig.~\ref{fig:Resonant Scatteirng schematic}, the calculations can be significantly simplified. In this resonant approximation, $\omega_i~\approx~E_\nu~-~E_i$, the scattering amplitude takes the form
\begin{equation}
\label{Eq.: Scattering matrix element resonant}
    \mathcal{M}^{res}_{M_f,M_i}=r_0 m c^2 \sum_{M_\nu}\frac{\braket{f|\hat{\mathcal{R}}^\dagger\left(\bm{k}_f,\bm{\epsilon}_f\right)|\nu}\braket{\nu|\hat{\mathcal{R}}\left(\bm{k}_i,\bm{\epsilon}_i\right)|i}}{E_i-E_\nu+\omega_i+i\Gamma_\nu/2},
\end{equation}
with $\ket{i}=\ket{\xi_i J_i M_i}$, $\ket{\nu}=\ket{\xi_\nu J_\nu M_\nu}$, and $\ket{f}=\ket{\xi_f J_f M_f}$ denoting the initial, intermediate, and final ionic states. These states are defined by their total angular momentum $J$, its projection $M$, and $\xi$ which accounts for all other quantum numbers necessary for a unique identification of the states. Moreover, $\Gamma_\nu$ is the natural width of the intermediate state and $r_0$ and $m$ are the classical electron radius and electron mass. The sum in Eq.~\eqref{Eq.: Scattering matrix element resonant} runs over the degenerate sublevels of the excited state with different angular momentum projections $M_\nu$~\cite{Samoilenko2020, Serbo22, Volotka22}.
The photon absorption and emission operators, $\hat{\mathcal{R}}$ and $\hat{\mathcal{R}}^\dagger$, can be expressed as a sum of their single-particle counterparts, written here in Coulomb gauge,
\begin{equation}
\label{eq: photon operator}
    \hat{\mathcal{R}}\left(\bm k, \bm \epsilon\right)=\sum_q \bm\alpha_q \cdot \bm\epsilon e^{i \bm{k}\bm{r}_q},
\end{equation}
where the index $q$ refers to the $q$-th electron.
Here, $\bm\alpha$ is the vector of Dirac matrices and $\bm{\epsilon}$ and $\bm{k}$ are the photon polarization and momentum vectors, respectively.
The matrix elements of these operators are usually evaluated by using a multipole expansion and the Wigner-Eckart theorem~\cite{Serbo22,Volotka22,Samoilenko2020}. \\
Amplitude~\eqref{Eq.: Scattering matrix element resonant} was frequently used to describe the process of resonant elastic photon scattering in the field-free case~\cite{Serbo22,Volotka22,Samoilenko2020}. 
It can be extended, however, for a target atom exposed to external electric and magnetic fields~\cite{Stenflo1998}. The effect of these external fields can also be treated within perturbation theory, leading to the modification of the ionic levels:
\begin{subequations}
\label{eq: modified energy levels}
\begin{align}
    E_i &= E_i^{(0)}+\Delta E_{Z,i}+\Delta E_{S,i} \label{eq. initial state energy}\\
    E_\nu &= E_\nu^{(0)}+\Delta E_{Z,\nu}+\Delta E_{S,\nu}  \label{eq. intermediate state energy}
\end{align}
\end{subequations}
and generally of the ionic wave functions. These wave functions and energies can be used to describe the coupling to the incoming laser field which again can be treated perturbatively. In this approach, the amplitude can still be written in the form of Eq.~\eqref{Eq.: Scattering matrix element resonant} where the energies are given by Eq.~\eqref{eq: modified energy levels}.

One has to note here that the influence of the external fields on the wave functions is neglected in the present work. Indeed, the external electric field also leads to the mixing of opposite parity states which results in additional interference effects~\cite{Richter2022}.  However, as we focus in this study on $J_i=0 \to J_\nu=1 \to J_f=0$ electric dipole transitions, the electric field induced state mixing leads to additional magnetic dipole transitions which are generally much smaller and hence can be neglected. 
\subsection{Cross Section and Polarization}
\label{sec: density matrix}
With the help of the scattering amplitude~\eqref{Eq.: Scattering matrix element resonant} and the energies \eqref{eq. initial state energy}-\eqref{eq. intermediate state energy}, one can evaluate the basic properties of the scattering process, such as the cross section and polarization of the emitted photons. Most naturally, this can be done within the framework of the density matrix approach~\cite{Blum2012,Samoilenko2020,Volotka22}. Within this theory the density matrix of scattered photons can be expressed as 
% In the helicity representation of the photon states, the matrix form of Eq.~\eqref{Eq.: Statistical Operator} for the scattering process is given by
\begin{align}
    &\braket{\bm{k}_f, \lambda_f | \hat{\rho}_f | \bm{k}_f, \tilde{\lambda}_f}\notag\\
    &=\frac{1}{2 J_i +1}\sum_{\lambda_i \tilde{\lambda}_i}\sum_{M_i M_f}\braket{\bm{k}_i, \lambda_i | \hat{\rho}_i | \bm{k}_i, \tilde{\lambda}_i}\notag \\
    & \times \mathcal{M}^{res}_{M_f,M_i}\left(\lambda_i,\lambda_f\right) \mathcal{M}^{res *}_{M_f,M_i}\left(\tilde{\lambda}_i,\tilde{\lambda}_f\right). 
    \label{Eq.: density matrix elements}
\end{align}
Here, $\mathcal{M}^{res}_{M_f,M_i}\left(\lambda_i,\lambda_f\right)$ is the scattering amplitude~\eqref{Eq.: Scattering matrix element resonant} written for a special case of circularly polarized incoming and outgoing photons with the helicity $\lambda_{i,f}$, and $\braket{\bm{k}_i, \lambda_i | \hat{\rho}_i | \bm{k}_i, \tilde{\lambda}_i}$ is the density matrix of the incident photons. Moreover in Eq.~\eqref{Eq.: density matrix elements} the target atoms are assumed to be unpolarized.
By making use of the photon density matrix, one can easily access the parameters of the scattering process as for instance the differential cross section:
\begin{equation}
    \frac{\mathrm{d}\sigma}{\mathrm{d}\Omega}\left(\theta_f,\phi_f,\omega_i\right) = \sum_{\lambda_f} \braket{\bm{k}_f \lambda_f| \hat{\rho}_f | \bm{k}_f \lambda_f}.
     \label{Eq.: density matrix differential cross section}
\end{equation}
% where $\lambda_{i,f}$ describes the helicity of the incoming and outgoing photon and the weight function $W\left(M_i\right)$ represents the population of the magnetic sublevels of the initial atomic state. In the case of unpolarized atoms, the weight function is given by  $W\left(M_i\right)=1/\left(2 J_i +1\right)$. In order to analyze the polarization of incoming and outgoing light, the photon density matrices can be expressed by the three Stokes parameters
Moreover, the polarization of the photons can be naturally expressed in terms of the three Stokes parameters which can be related to the density matrix by
\begin{equation}
  \braket{\bm{k}\lambda|\hat{\rho}|\bm{k}\tilde{\lambda}} =  \mathcal{N} \left( \begin{matrix} 1+P_3 & -P_1+iP_2 \\ -P_1-iP_2 & 1-P_3 \end{matrix} \right),
  \label{Eq.: density matrix}
\end{equation}
where $\mathcal{N}=1$ describes the incoming radiation and $\mathcal{N}=\mathrm{d}\sigma/\mathrm{d}\Omega$ the outgoing radiation. 
Here, the first two Stokes parameters describe linear polarization, either within the scattering plane or perpendicular to it, in the case of $P_1$, or at angles of $45^\circ$ or $135^\circ$, in the case of $P_2$. The parameter $P_3$ characterizes circular polarization.

\section{Scattering off He-like ions in External Fields}\label{Section:He-like Ions}
The theory introduced in the previous sections is general and can be applied to any ion. In what follows, we will consider the particular case of $1\text{s}^2$ $^1\text{S}_0$ $ \to$ $1\text{s}2\text{p}$ $^1\text{P}_1 $ scattering of photons by He-like Ca ions. This process will be discussed for typical parameters of the Gamma Factory project at CERN. In particular, in the analysis below we assume the Lorentz factor of $\gamma_L\approx 2395$, initial photons with energy $\omega_\mathrm{lab}\approx 0.815$\,eV which can be delivered by commercially available erbium-doped fiber lasers~\cite{Nichita:2021iwa}
% \footnote{Photons of such energy can be delivered by the commercially available erbium-doped fiber laser, see e.g https://www.ipgphotonics.com/.}
and external magnetic field strengths up to $B_\mathrm{lab}=3$\,T.

\subsection{Energy Shifts of the $1\text{s}^2$ $^1\text{S}_0$ and $1\text{s}2\text{p}$ $^1\text{P}_1 $ States}
\begin{figure}
    \centering
    \includegraphics[width=0.5\textwidth]{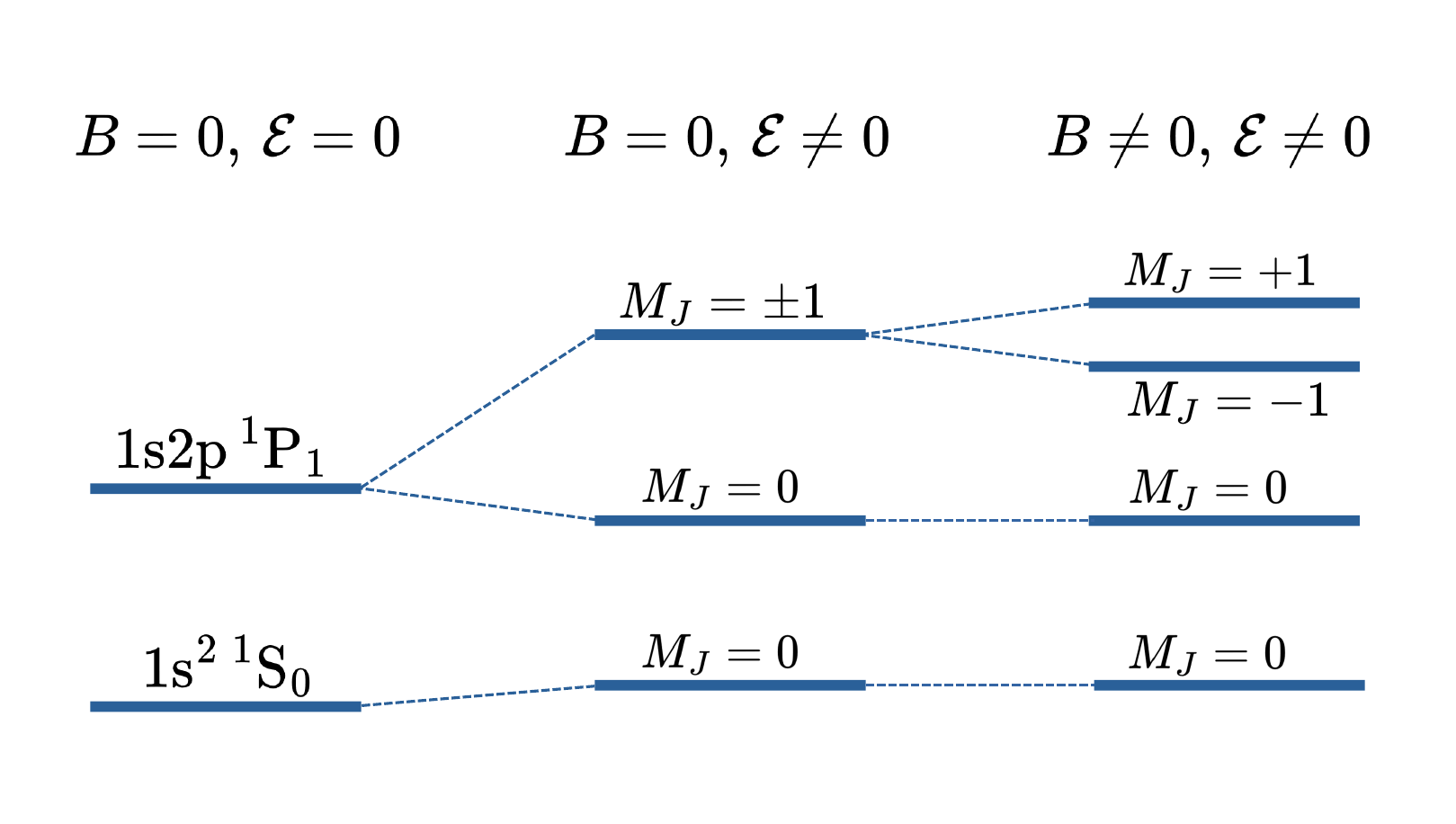}
    \caption{Schematic illustration of the Zeeman and Stark shift on the $1\text{s}^2$ $^1\text{S}_0$ and $1\text{s}2\text{p}$ $^1\text{P}_1 $ states.}
    \label{fig: energy shift schematic}
\end{figure}

As discussed already in Sections~\ref{Section:EnergyShifts} and \ref{Section:ResonantScattering}, the splitting and shift of ionic levels due to Zeeman and Stark effects may affect the scattering process and hence have to be well understood. In Fig.~\ref{fig: energy shift schematic}, therefore, we present a schematic illustration of how the
$1\text{s}^2$~$^1\text{S}_0$ and $1\text{s}2\text{p}$~$^1\text{P}_1 $ magnetic sublevels are affected by external electric and magnetic fields in the ion rest frame.
As seen in this figure, the Stark shift $(\mathcal{E}\neq 0)$ leads to a shift of the $\ket{^1\text{S}_0,\, M_i=0}$ sublevel and a splitting of the $\ket{^1\text{P}_1,\, M_\nu=0}$ and $\ket{^1\text{P}_1,\, M_\nu=\pm 1}$ sublevels. This prediction can be well understood based on Eq.~\eqref{Eq.: Stark Shift}, which predicts that the quadratic Stark shift depends on $M_J^2$.
If in addition to the electric field an external magnetic field is applied, the ionic levels also exhibit first and second order Zeeman shift.   
While the second order Zeeman shift is usually small and depends on $M_J^2$, the first order correction is linear in $M_J$ and leads to a splitting of the $\ket{^1\text{P}_1,\, M_\nu=+ 1}$ and $\ket{^1\text{P}_1,\, M_\nu=- 1}$ sublevels. The combined effect of electric and magnetic fields on the magnetic sublevels is displayed in the right column of Fig.~\ref{fig: energy shift schematic}.

So far we have only presented the qualitative picture of the splitting 
of $^1\text{S}_0$ and $^1\text{P}_1$ sublevels for ions exposed to external electromagnetic fields. 
Using the equations discussed in the previous sections, one can also derive formulas for a quantitative analysis. 
Indeed, by combining Eqs. \eqref{Eq.: Lorentz transformation fields}, \eqref{Eq.: Zeeman first order}, \eqref{Eq.: Stark Shift} and~\eqref{eq: modified energy levels} we obtain
\begin{subequations}
\label{Eq.: Zeeman and Stark shift 1S0 and 1P1}
\begin{align}
      &\Delta E_\mathrm{tot} \left(^1\text{S}_0\right) = -\frac{1}{2} \alpha_0\left(^1\text{S}_0\right) \gamma_L^2 v^2 B_{\mathrm{lab}}^2,\\
      &\Delta E_\mathrm{tot} \left(^1\text{P}_1\right) =\begin{aligned}[t]
          &-\frac{1}{2} \alpha_0\left(^1\text{P}_1\right) \gamma_L^2 v^2 B_{\mathrm{lab}}^2\\
      & +\frac{3M_\nu^2-2}{4} \alpha_2\left(^1\text{P}_1\right) \gamma_L^2 v^2 B_{\mathrm{lab}}^2\\
      &+M_\nu g \mu_B \gamma_L B_{\mathrm{lab}},
      \end{aligned}
\end{align}
\end{subequations}
where $\Delta E_\mathrm{tot}=\Delta E_Z + \Delta E_S$ is the sum of Zeeman and Stark shifts, and $\alpha_0$ and $\alpha_2$ are the scalar and tensor polarizabilities.
Moreover in Eq.~\eqref{Eq.: Zeeman and Stark shift 1S0 and 1P1}, the second order Zeeman effect is neglected since, according to our calculations, it is much smaller than $\Delta E_Z^{(1)}$. Indeed, for Ca$^{18+}$ ions moving with $\gamma_L=2395$ in the presence of the external magnetic field with $B_\mathrm{lab}=1$\,T we found $\Delta E_Z^{(1)}\approx0.1$\,eV and $\Delta E_Z^{(2)}\approx10^{-5}$\,eV.

As seen from Eq.~\eqref{Eq.: Zeeman and Stark shift 1S0 and 1P1} the calculations of the energies of the $\ket{^1\text{S}_0}$ and $\ket{^1\text{P}_1,M_\nu}$ sublevels for the scenario of interest requires evaluations of the $g$-factor and the polarizabilities $\alpha_0$ and $\alpha_2$. While the $g$-factor of the $^1\text{P}_1$ state is given as $g\approx1$~\cite{Puchalski2012}, the polarizabilities are calculated using the configuration interaction method implemented in the AMBiT code, thoroughly explained in Ref.~\cite{KAHL2019}.
The polarizabilities and the transition energies for the unperturbed case as well as the lifetimes of the excited $^1\text{P}_1$ state are shown in Table~\ref{tab:Polarizabilities}. 
By making use of these data and Eq.~\eqref{Eq.: Zeeman and Stark shift 1S0 and 1P1} one can finally calculate the Zeeman and Stark shift of the ground and excited ionic states. This calculation indicates that $\Delta E_\mathrm{tot}\left( ^1\text{S}_0 \right)$ is much smaller than $\Delta E_\mathrm{tot}\left( ^1\text{P}_1 \right)$. 
In Table~\ref{tab:Energy shifts} therefore, we present the energy shifts for the excited sublevels $\ket{^1\text{P}_1, M_\nu}$ only. From this table one can conclude that the Stark shift is the dominant effect for the chosen experimental parameters and He-like Ca projectiles.

{\renewcommand{\arraystretch}{1.4}
\begin{table*}[t]
    \centering
    \begin{ruledtabular}
    \begin{tabular}{llllll}
          & $E^{(0)}_{^1P_1}-E^{(0)}_{^1S_0} $ in eV&$\alpha_0\left(^1S_0\right)$ in a.u.&$\alpha_0\left(^1P_1\right)$ in a.u.&$\alpha_2\left(^1P_1\right)$ in a.u.& $\tau\left(^1P_1\right)$ in s\\
          \hline 
        He  & $21.212$ &  $1.36$ & $-5.99 \times 10^1$& $2.23 \times 10^2$ &$5.57 \times 10^{-10}$\\
        &$21.218$~\cite{Pachuki2017} &$1.38$~\cite{Puchalski2016} & $-6.00 \times 10^1$~\cite{Yan2000} &$2.24  \times 10^2$~\cite{Bhaskar1974} & $5.51 \times 10^{-10}$~\cite{Notermans2014}\\
                &&&&&\\
        $\text{Ca}^{18+}$  &$39.039 \times 10^2$ & $5.67 \times 10^{-5}$& $-2.27 \times 10^{-2}$& $2.38 \times 10^{-2}$& $6.11 \times 10^{-15}$ \\
        &$39.024 \times 10^2$~\cite{Yerokhin2019}&&&&$6.09 \times 10^{-15}$~\cite{Si2016}\\
                % &&&&&\\
        % $\text{Xe}^{52+}$  & $30.674 \times 10^3$& $8.41 \times 10^{-7}$& $-1.03 \times 10^{-4}$& $1.05 \times 10^{-4}$ & $1.46 \times 10^{-16}$\\
        % &$30.630 \times 10^3$~\cite{Yerokhin2019}&&&&\\
    \end{tabular}
    \end{ruledtabular}
    \caption{The $1\text{s}^2$ $^1\text{S}_0$ $ \to$ $1\text{s}2\text{p}$ $^1\text{P}_1 $ transition energies, the scalar and tensor polarizabilities (in atomic units) and the lifetimes of the $^1\text{P}_1$ state for neutral He and He-like Ca. Our results, based on CI calculations using the AMBiT code~\cite{KAHL2019},  agree well with available data from Ref.~\cite{Pachuki2017,Puchalski2016,Yan2000,Notermans2014,Yerokhin2019,Si2016}.  }
    \label{tab:Polarizabilities}
\end{table*}
}

% {\renewcommand{\arraystretch}{1.4}
% \begin{table*}[t]
%     \centering
%     \begin{ruledtabular}
%     \begin{tabular}{llll|lll}
%          & \multicolumn{3}{c}{$M_J = 0$} & \multicolumn{3}{c}{$M_J = \pm 1$}    \\ \hline
%          & $\Delta{E_S}$ in eV& $\Delta{E_Z^{(1)}}$ in eV& $\Delta{E_Z^{(2)}}$ in eV& \multicolumn{1}{l}{$\Delta{E_S}$ in eV} & $\Delta{E_Z^{(1)}}$ in eV& $\Delta{E_Z^{(2)}}$ in eV \\
%          $\text{Ca}^{18+}$  & $-2.92 \times 10^{-2}$& $0$ &$2.3 \times 10^{-5}$ & \multicolumn{1}{l}{$9.18\times10^{-1}$} & $\pm1.39\times10^{-1}$& $1.3 \times 10^{-5}$\\
%         % $\text{Xe}^{52+}$ & $-5.31 \times 10^{-5}$& $0$ & $5.0 \times 10^{-5}$ & \multicolumn{1}{l}{$4.12 \times 10^{-3}$} & $\pm1.39\times10^{-1}$&$3.3 \times 10^{-5}$ \\ 
%     \end{tabular}
%     \end{ruledtabular}
%     \caption{Resulting Zeeman and Stark shifts of the $1\text{s}2\text{p}$ $^1\text{P}_1 $ state in He-like Ca and He-like Xe for a magnetic field strength in the laboratory frame of $B_{\mathrm{lab}}=1$ T and a Lorentz factor of $\gamma=2395$.}
%     \label{tab:Energy shifts}
% \end{table*}
% }
{\renewcommand{\arraystretch}{1.4}
\begin{table}[t]
    \centering
    \begin{ruledtabular}
    \begin{tabular}{rrrr}
        
        $M_J$ & $\Delta{E_S}$ in eV& $\Delta{E_Z^{(1)}}$ in eV& $\Delta{E_\mathrm{tot}}$ in eV \\ \hline
         $0$  & $-2.92 \times 10^{-2}$& $0$ &$-2.92 \times 10^{-2}$ \\
         $1$& \multicolumn{1}{l}{$9.18\times10^{-1}$} & $1.39\times10^{-1}$& $10.57 \times 10^{-1}$\\
        $-1$& \multicolumn{1}{l}{$9.18\times10^{-1}$} & $-1.39\times10^{-1}$& $7.79 \times 10^{-1}$
         
    \end{tabular}
    \end{ruledtabular}
    \caption{Resulting Zeeman and Stark shifts of the $1\text{s}2\text{p}$ $^1\text{P}_1 $ state in He-like Ca for a magnetic field strength in the laboratory frame of $B_{\mathrm{lab}}=1$ T and a Lorentz factor of $\gamma_L=2395$.}
    \label{tab:Energy shifts}
\end{table}
}

\subsection{External Field Effects on Resonant Scattering}

In the previous subsection, we have briefly discussed how external electric and magnetic fields modify the sublevels $\ket{^1\text{S}_0, M_i}$ and $\ket{^1\text{P}_1, M_\nu}$ of He-like Ca ions. 
Now we are ready to explore, how such fields affect the properties of the $^1\text{S}_0 \to $ $^1\text{P}_1$ resonant photon scattering.
In the next subsection, for instance, we will discuss the scattering rate of photons as observed in the foreseen Gamma Factory setup.
Their angular distribution and polarization for the case of initially circularly polarized light will be studied then in Subsection~\ref{sec: angular distribution and polarization}.
The presented formalism can be applied to an arbitrary polarization of the laser photons. 
The choice of circularly polarized photons is related to their important role for the Gamma Factory project, e.g. for investigations of atomic parity violation~\cite{Richter2022} or the production of longitudinally polarized muon and positron beams~\cite{Apyan:2022ysh}.

\subsubsection{Rate of Scattered Photons}\label{Section: Cross section He-like Ca}
\begin{figure}   
    \includegraphics[width=0.45\textwidth]{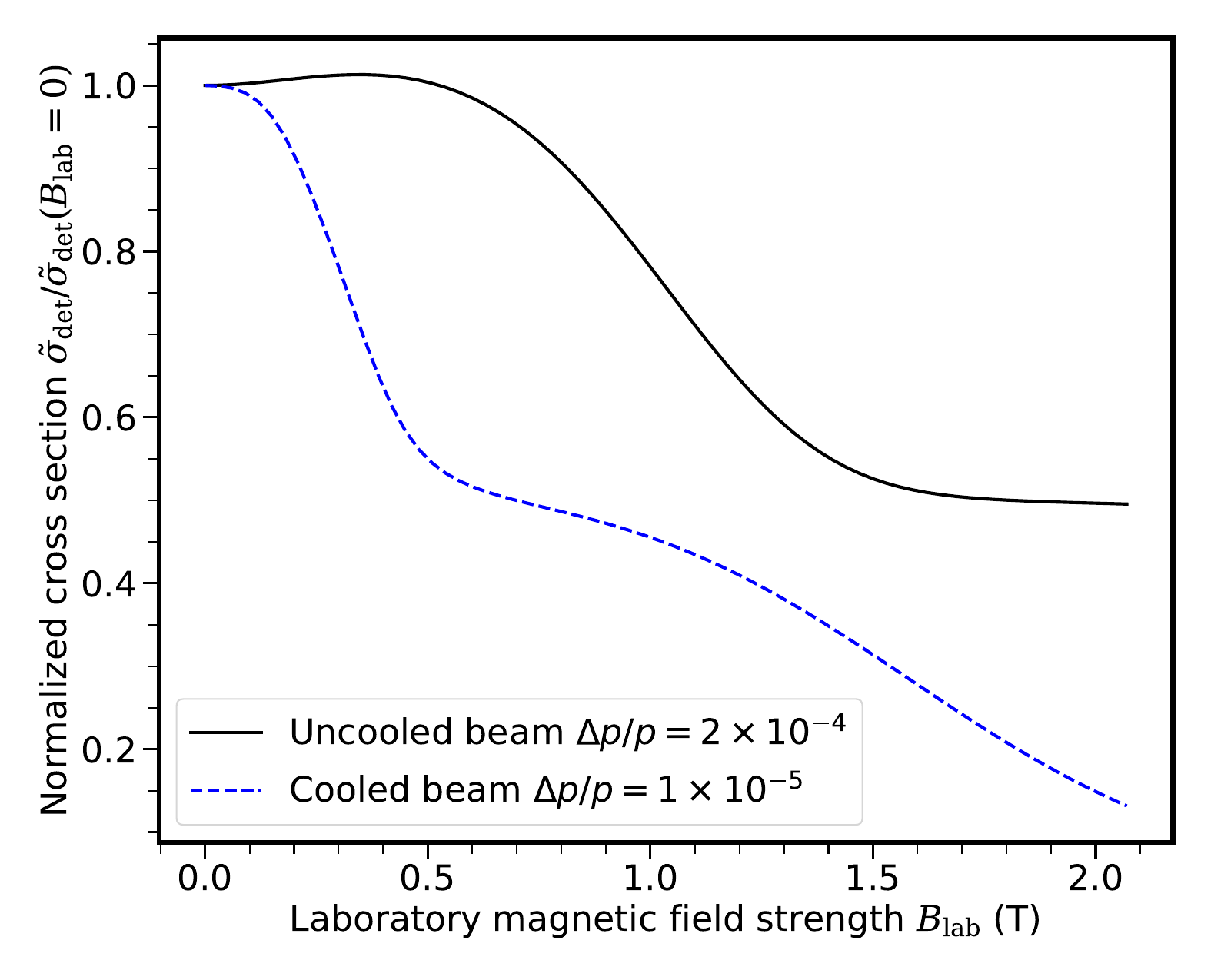}
    \caption{The normalized cross section of photons scattered in a finite solid angle in y-direction with an opening angle of 1\,mrad in the laboratory frame for He-like Ca ions. The cross section is shown as a function of the magnetic field strength in the laboratory frame. The black solid line is calculated for the uncooled beam scenario, $\Delta p/p = 2\times 10^{-4}$, and the blue dashed line represents the cooled beam scenario,  $\Delta p/p = 1 \times 10^{-5}$.}
    \label{fig:Total Cross section plot}
\end{figure}

Photons resonantly scattered by ultra-relativistic projectiles are emitted in the laboratory frame within a small solid angle around the ion beam propagation direction.
For the typical parameters of the Gamma Factory, for example, the majority of the photons will be scattered within an opening angle $\theta_f^{(\mathrm{lab})}\leq 1$\,mrad. 
Therefore, detectors covering a comparable solid angle range will detect most of the scattered radiation.
In order to investigate, how the count rate observed by such detectors will depend on the strength of the laboratory magnetic field, we have performed calculations, whose results are summarized in Fig.~\ref{fig:Total Cross section plot}.
In this figure, we display the differential cross section, integrated over the solid angle of the detector ($\theta_f^{(\mathrm{lab})}\leq 1$\,mrad),
\begin{equation}
\label{eq.: cross section detector}
\sigma_\mathrm{det} \left(\omega_i, B_\mathrm{lab}\right)= \int_{\Omega_\mathrm{det}} \frac{\mathrm{d}\sigma}{\mathrm{d}\Omega_f}\left(\theta_f,\phi_f,\omega_i, B_\mathrm{lab}\right) \,\mathrm{d}\Omega_f,    
\end{equation}
normalized to the predictions for $B_\mathrm{lab}=0$. 
For the latter case and for zero detuning, the integrated cross section~\eqref{eq.: cross section detector} can be estimated as $\sigma_\mathrm{det}\left(\omega_i=E_\nu-E_i,B_\mathrm{lab}=0\right)\approx 4\times 10^8$\,barn.
Moreover, we have taken into account the momentum spread of the ion beam, which leads to the fact that the laser radiation, which is assumed to be monochromatic in the laboratory frame, is Doppler broadened in the ion frame.
To account for this broadening, the cross section is convoluted with a Gaussian frequency distribution:
\begin{equation}
\label{eq: gaussian average}
    \tilde\sigma_\mathrm{det}\left(B_\mathrm{lab}\right) = \int \sigma_\mathrm{det}\left(\omega_i,B_\mathrm{lab}\right) \frac{1}{ \sqrt{2 \pi \Delta\omega^2}} e^{-\frac{(\omega - \omega_\mathrm{ion})^2}{2 \Delta\omega^2}} \, \mathrm{d}\omega_i, 
\end{equation}
where $\omega_\mathrm{ion}= 2 \gamma_L \omega_\mathrm{lab}$ is the mean laser frequency in the ion frame, and $\Delta \omega$ is its width due to the ion beam’s momentum spread. 
As mentioned already in Sec.~\ref{Section:Geometry}, scenarios of uncooled and cooled ion beams are considered, corresponding to relative widths of $\Delta\omega/\omega_\mathrm{ion}=2\times10^{-4}$ and $\Delta\omega/\omega_\mathrm{ion}=1\times 10^{-5}$, respectively.
The results for these scenarios are depicted by the solid and dashed lines in Fig.~\ref{fig:Total Cross section plot}.
As seen from this figure, the normalized rate of detected photons is very sensitive to the laboratory magnetic field strength. 
The effect of the magnetic field is strongest for the cooled ion beam, where the normalized rate falls to approximately $\tilde{\sigma}_\mathrm{det}(B_\mathrm{lab})/\tilde{\sigma}_\mathrm{det}(0)~\approx~0.15$ as $B_\mathrm{lab}$ increases from zero to $2$\,T. In the uncooled beam scenario, $\tilde{\sigma}_\mathrm{det}(B_\mathrm{lab})/\tilde{\sigma}_\mathrm{det}(0)$ also declines with increasing $B_\mathrm{lab}$, though this effect is less pronounced. Additionally, for the uncooled beam, the count rate stabilizes around $\tilde{\sigma}_\mathrm{det}(B_\mathrm{lab})/\tilde{\sigma}_\mathrm{det}(0)~\approx~0.5$ for $B_\mathrm{lab} > 1.5$\,T.

\begin{figure}   
    \includegraphics[width=0.45\textwidth]{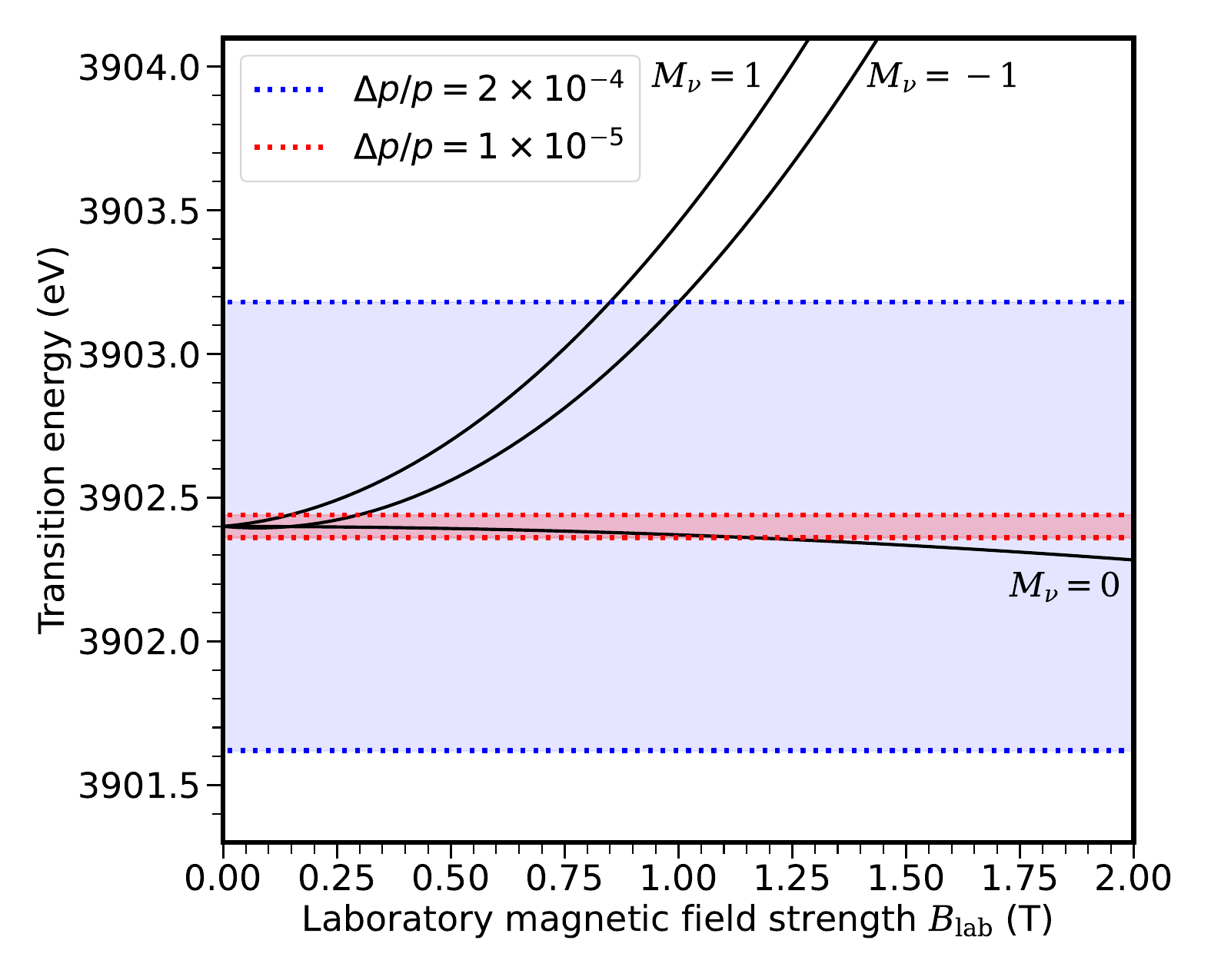}
    \caption{The transition energies from the ground state to the magnetic $\ket{^1\text{P}_1, M_\nu}$ sublevels as a function of the laboratory magnetic field strength. The red and blue dotted lines represent the frequency width of the incoming radiation in the ion frame for both scenarios of uncooled (blue) and cooled (red) ion beams. The natural line width $\Gamma_\nu=\hbar/\tau$ is neglected in this visualization.}
    \label{fig:Transition energy plot}
\end{figure}

In order to gain a qualitative understanding of this behaviour, we inspect Fig.~\ref{fig:Transition energy plot}. This figure, shows the transition energies from the ground state to the different magnetic sublevels $\ket{^1\text{P}_1, M_\nu}$ as a function of the magnetic field strength, compared to the width of the frequency distribution of the incident radiation as seen in the ion frame.
This figure reveals that $M_\nu=\pm 1$ sublevels are quickly shifted out of resonance in the cooled beam scenario, leading to the rapid drop of the rate of detected photons. For the uncooled beam,  however, the rate begins to decrease not until $B_\mathrm{lab}>0.5$\,T because of the larger frequency width.  
Moreover, the $M_\nu = 0$ sublevel stays in resonance at higher $B_\mathrm{lab}$ values in the uncooled beam scenario, leading to the observed stagnation of the count rate at $\tilde{\sigma}_\mathrm{det}(B_\mathrm{lab})/\tilde{\sigma}_\mathrm{det}(0) \approx 0.5$.\\ 

The qualitative understanding gained from Fig.~\ref{fig:Transition energy plot} can be extended to a quantitative analysis by 
studying the relative contributions of the different $\ket{^1\text{P}_1, M_\nu}$ sublevels to the scattering process. 
By using the photon absorption operator~\eqref{eq: photon operator} and some angular momentum algebra \cite{balashov2013, rose1957}, it can be shown that the photo-excitation rate from the $\ket{^1\text{S}_1,M_i=0}$ ground state to one excited $\ket{^1\text{P}_1, M_\nu}$ sublevel is proportional to the absolute square of the small Wigner-$d$ matrix:
\begin{align}
    W_{\ket{^1\text{S}_0,M_i=0} \to \ket{^1 \text{P} _ 1, M_\nu}}
    \propto \left|d^1_{M_\nu\lambda}\left(\theta_i-\frac{\pi}{2}\right)\right|^2 \left|a_{E1}\right|^2.
    \label{eq: Excitation rate}
\end{align}
Here, $a_{E1}$ is the reduced transition matrix element and $\theta_i~-~\pi/2~=~\pi/2$ is the angle between the initial photon propagation direction and the quantization axis. 
From Eq.~\eqref{eq: Excitation rate}, it follows immediately that the $M_\nu=\pm1$ sublevels each contribute $25\%$ to the total excitation rate, while the remaining $50\%$ are provided by the $M_\nu~=~0$ sublevel. Based on this observation, one can easily understand why the rate of detected photons reaches a steady value around $\tilde{\sigma}_\mathrm{det}(B_\mathrm{lab})/\tilde{\sigma}_\mathrm{det}(0) \approx 0.5$ for high magnetic fields in the uncooled beam scenario.

\begin{figure}   
    \includegraphics[width=0.45\textwidth]{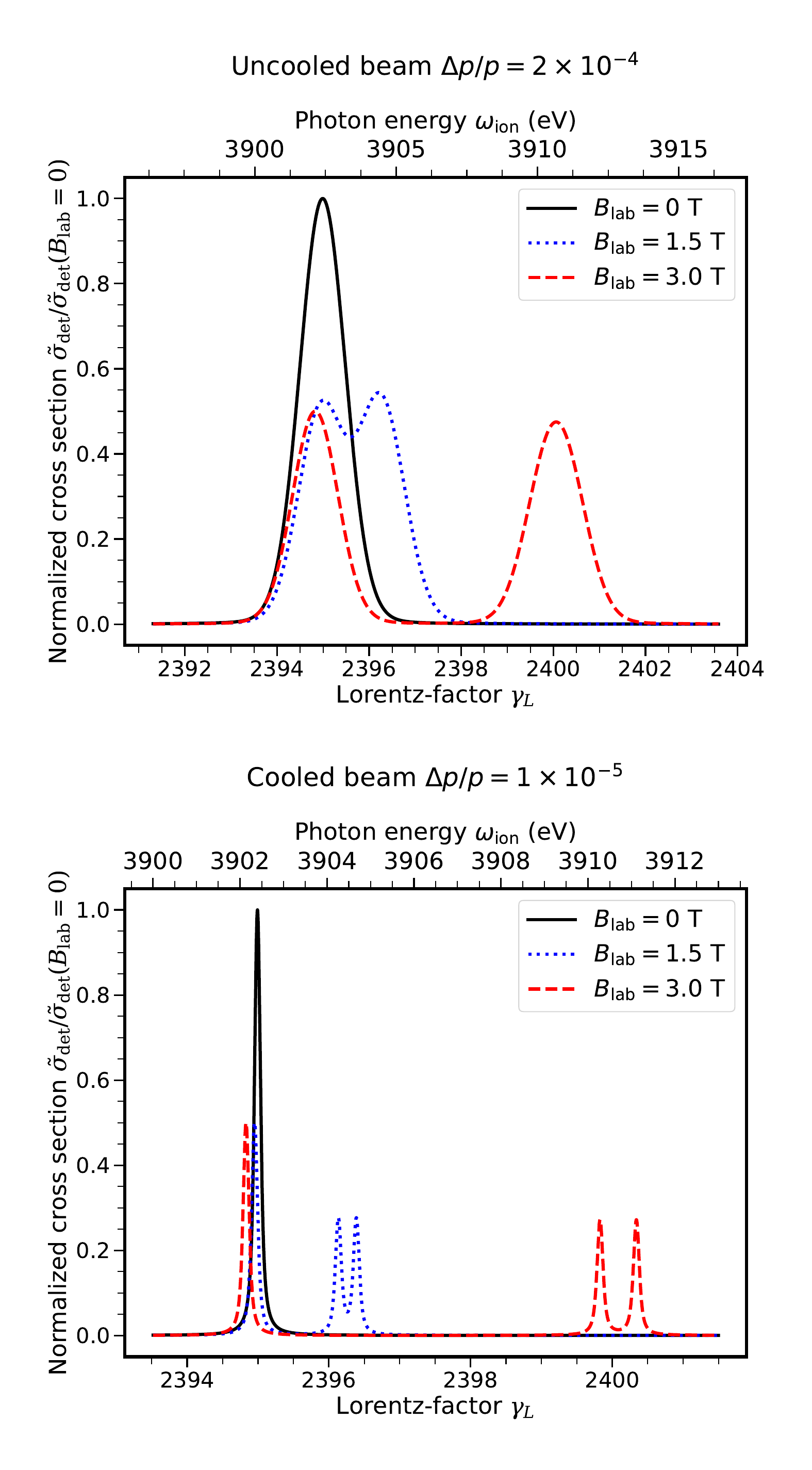}
    \caption{The normalized cross section of photons scattered in a finite solid angle in y-direction with an opening angle of 1\,mrad in the laboratory frame for He-like Ca ions. The cross section is shown as a function of the Lorentz factor $\gamma_L$ and the corresponding laser photon energy as seen in the ion frame. The upper plot is calculated for the uncooled beam scenario, $\Delta p/p = 2\times 10^{-4}$, and the lower plot represents the cooled beam scenario, $\Delta p/p=1 \times 10^{-5}$.}
    \label{fig:Total Cross section over omega plot}
\end{figure}

So far, our analysis was focused on a particular Lorentz factor $\gamma_L=2395$. We now extend this study to a variable Lorentz factor, as it is used in the Gamma Factory setup to scan over the photon frequency in the ion frame. Therefore, we calculate the normalized countrate of detected photons as a function of the Lorentz factor by varying $\omega_\mathrm{ion}=2 \gamma_L\omega_\mathrm{lab}$ in Eq.~\eqref{eq: gaussian average}. The resulting spectrum, as it could be observed in a Gamma Factory experiment, is shown in Fig.~\ref{fig:Total Cross section over omega plot} for three different laboratory magnetic field strength.
This figure demonstrates that the Stark splitting between the $\ket{^1\text{P}_1, M_\nu=0}$ and $\ket{^1\text{P}_1, M_\nu=\pm1}$ sublevels can be observed in both uncooled and cooled beam settings, whereas the smaller Zeeman splitting of the $\ket{^1\text{P}_1, M_\nu=\pm1}$ sublevels is resolved only in the cooled beam scenario.

\subsubsection{Angular Distribution and Polarization}
\label{sec: angular distribution and polarization}
Alongside the modified rate of detected photons, which was discussed in the previous subsection, a laboratory magnetic field can have a notable impact on the angular and polarization properties of scattered radiation.
This impact of the $B$-field can also be attributed to the mutual shift of the magnetic sublevels $\ket{^1\text{P}_1, M_\nu}$ and is related to the well-known Hanle effect~\cite{hanle1924,Avan1975,Togawa2024,richter2024}.

\begin{figure*}[t]
    \centering
    \includegraphics[width=0.95\textwidth]{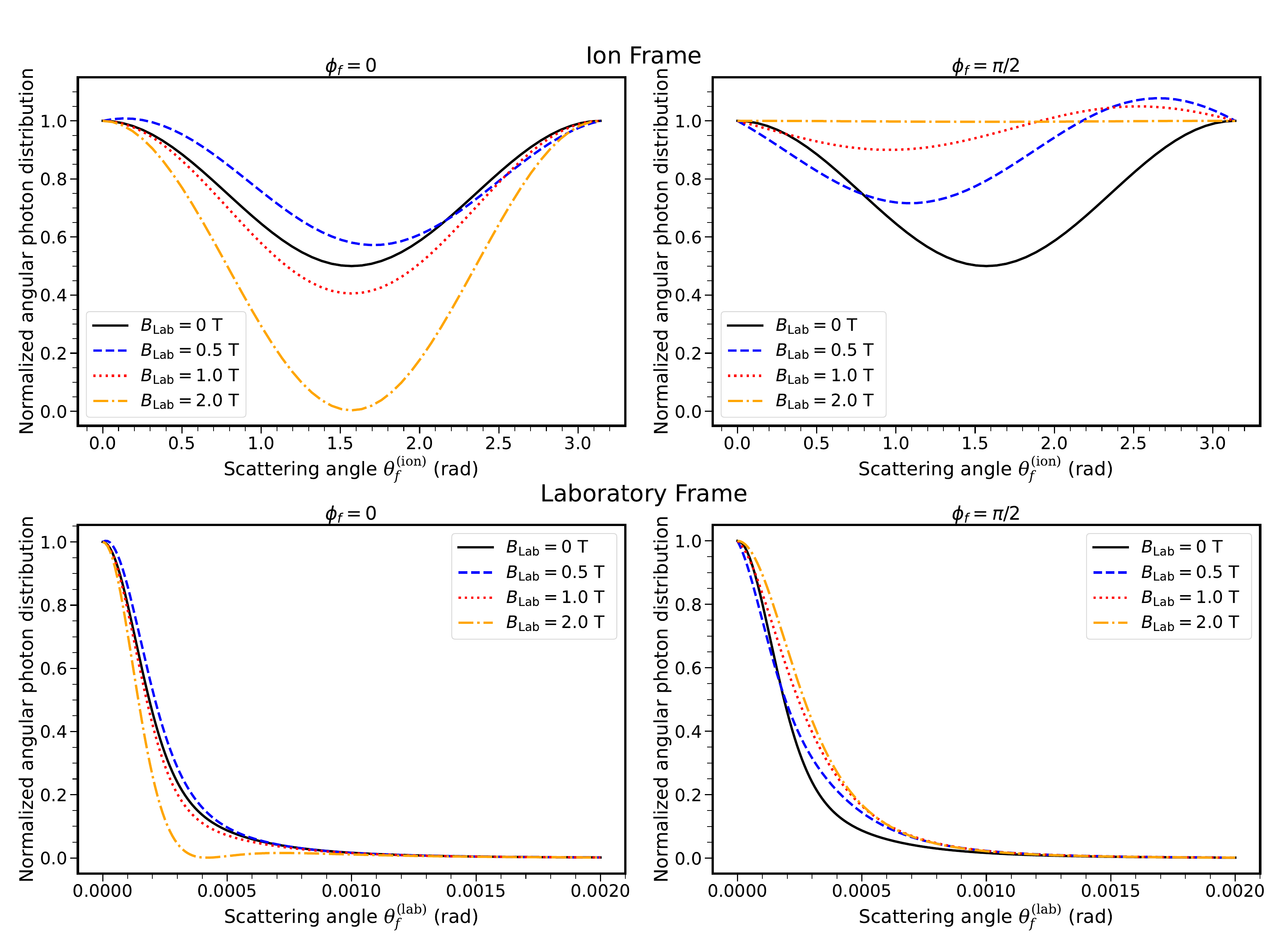}
    \caption{The angular distribution of the emitted photons in the $1\text{s}^2$ $^1\text{S}_0$ $ \to$ $1\text{s}2\text{p}$ $^1\text{P}_1 $ scattering process in He-like Calcium for $\gamma_L=2395$ and different laboratory magnetic field strengths.
    For each value of $B_\mathrm{lab}$ the angular distribution is normalized by its value for $\theta_f=0$.
    The upper panels show the result in the rest frame of the ion while the lower panels show the Lorentz transformed result in the laboratory frame.}
    \label{fig: Angular distribution plot}
\end{figure*}

To discuss, how the laboratory magnetic field can influence the angular distribution of emitted photons, we remind first the well established results for the field-free case, $B_\mathrm{lab}=0$,
\begin{equation}
W\left(\theta_f^{(\mathrm{ion})},\phi_f^{(\mathrm{ion})},B_\mathrm{lab}=0\right) = \frac{W_0}{4}\left(3+\cos2\theta_f^{(\mathrm{ion})}\right).
    \label{Eq.: angular distribution B=0}
\end{equation}
This formula is derived in the ion rest frame, for the case of a $J_i=0 \to J_\nu=1 \to J_f=0$ transition and circularly polarized incident light~\cite{Volotka22}.
%In the absence of external fields, the angular distribution of the emitted photons in the ion frame for circularly polarized incoming light is given by a simple analytic expression~\cite{Volotka22}:
To apply Eq.~\eqref{Eq.: angular distribution B=0} to describe an experiment one would need to average it over the photon frequency distribution in the ion frame similar to what is done in Eq.~\eqref{eq: gaussian average}. However, since for the field-free case~\eqref{Eq.: angular distribution B=0}, all magnetic sublevels are degenerate and, hence, the energy dependence only appears in the prefactor $W_0$, the shape of the angular distribution is not affected by the frequency averaging.
As seen from Eq.~\eqref{Eq.: angular distribution B=0} and Fig.~\ref{fig: Angular distribution plot}, the angular distribution of emitted photons for the field-free case is independent on the azimuthal angle $\phi_f^{(\mathrm{ion})}$ and is symmetric with respect to $\theta_f^{(\mathrm{ion})}=\pi/2$.

In order to explore, how the angular distribution~\eqref{Eq.: angular distribution B=0} is altered if an external magnetic field is applied, we employ the matrix element~\eqref{Eq.: Scattering matrix element resonant}, the modified energies~\eqref{eq: modified energy levels} and the general expression for the angle-differential cross section~\eqref{Eq.: density matrix differential cross section}, averaged over the photon frequency distribution.
The resulting emission patterns in the ion frame, obtained for the uncooled beam scenario, are displayed in the upper panels of Fig.~\ref{fig: Angular distribution plot}.
One can see from this figure, that the angular distribution of emitted photons is very sensitive to $B_\mathrm{lab}$. In particular, $W\left(\theta_f^{(\mathrm{ion})},\phi_f^{(\mathrm{ion})} , B_\mathrm{lab}\neq0\right)$ has a pronounced dependence on the azimuthal angle $\phi_f^\mathrm{ion}$ that can be attributed to the fact that electric and magnetic fields, as seen by the ion, break the azimuthal symmetry of the system.
Moreover, for the case of $\phi_f^{(\mathrm{ion})}=\pi/2$ the original symmetry to the polar angle $\theta_f^{(\mathrm{ion})}=\pi/2$ is clearly broken already for smaller magnetic fields, $B_\mathrm{lab}\approx0.5$\,T.\\

It is interesting to note that the angular distribution of scattered photons responds differently to weak and strong magnetic fields.
Most clearly it can be seen for $\phi_f^{(\mathrm{ion})} = 0$. 
For this geometry, the emission pattern becomes more isotropic as the magnetic field strength increases from $B_\mathrm{lab} = 0$ to $B_\mathrm{lab} \approx 0.5$\,T.
Further enhancement of $B_\mathrm{lab}$, however, leads to a more pronounced anisotropy of $W$.
This irregular behaviour of the angular distribution can be attributed to an interplay of two different effects.
Indeed, for small $B_\mathrm{lab}$ the Stark and Zeeman shifts remain small compared to the width of the Doppler-broadened frequency distribution, but are comparable to the natural width $\Gamma_\nu$.
In this regime, the behaviour of the angular distribution is governed by the Hanle effect which reduces the anisotropy of $W$. 
In contrast, for $B_\mathrm{lab} \geq 1$\,T, the $\ket{^1\mathrm{P}_1, M_\nu=\pm 1}$ sublevels are shifted out of resonance, see Fig.~\ref{fig:Transition energy plot}.
Hence, the scattering proceeds via a single sublevel $\ket{^1\mathrm{P}_1, M_\nu=\pm 0}$ resulting in a strong asymmetry of the angular distribution.\\ 

\begin{figure*}   
    \includegraphics[width=0.99\textwidth]{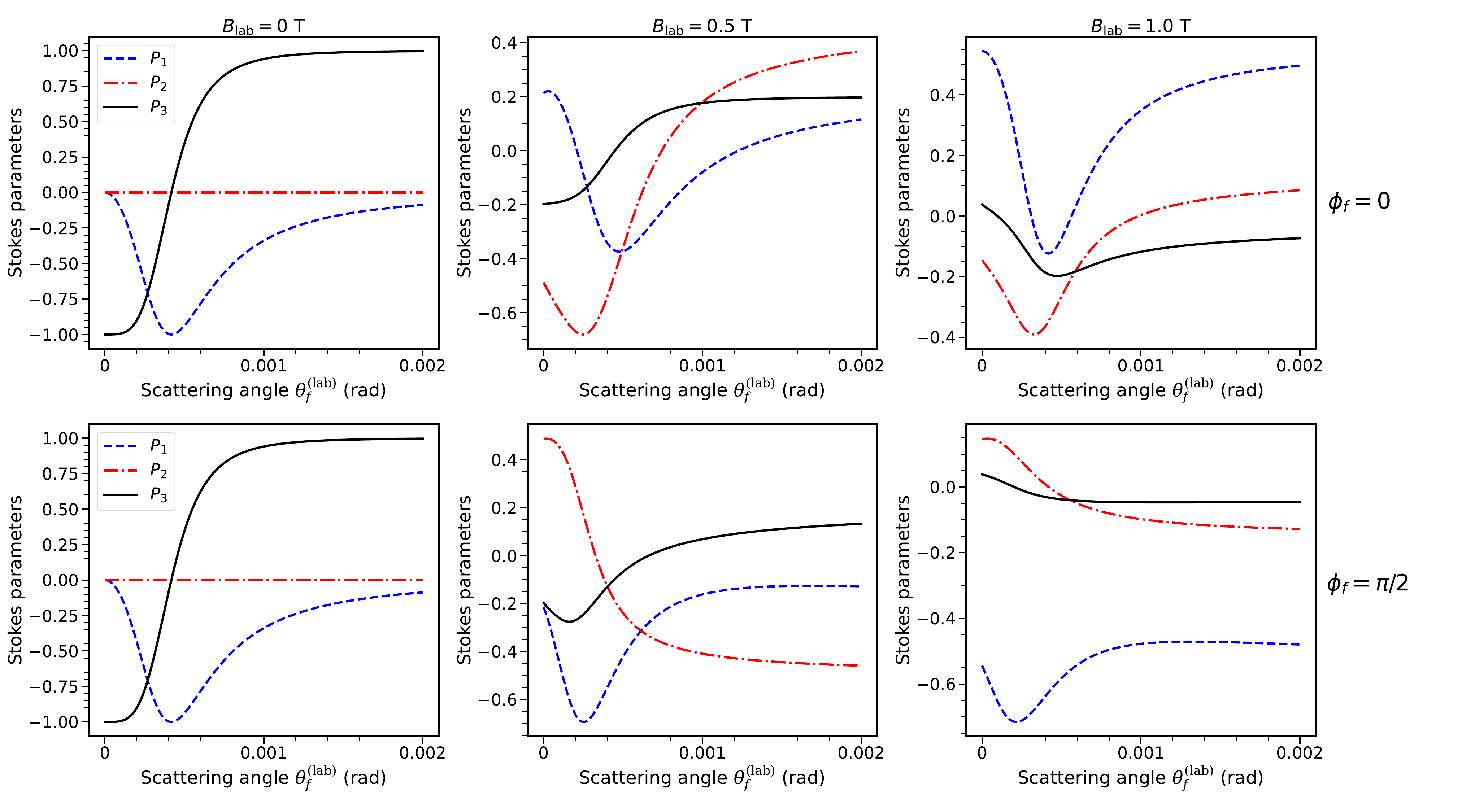}
    \caption{The three Stokes parameters $ P^{(f),(\mathrm{lab})}$ of the emitted photons as a function of the scattering angle in the laboratory frame for different magnetic field strengths.}
    \label{fig: Polarization Angular Plot}
\end{figure*}

So far we have only discussed the angular distribution of emitted photons in the ion frame.
However, since all measurements are performed in the laboratory frame, a transformation of our results to this frame is required, which can be easily executed by using Eqs.~\eqref{eq.: lorentz trafo angle} and \eqref{eq.: lorentz trafo solid angle}.
The resulting angular distributions of scattered photons in the laboratory frame are depicted in the lower panels of Fig.~\ref{fig: Angular distribution plot}.
This figure demonstrates the well known effect of a focusing of scattered radiation along the direction of the ion beam propagation, $\theta_f^{(\mathrm{lab})}\leq 1$\,mrad, due to the ultra-relativistic movement of the ions.
Despite of this focusing the influence of the external magnetic field can be observed even in the laboratory frame. In particular in the range $0.3$\,mrad$\,\leq\theta_f^{(\mathrm{lab})}\leq 0.6$\,mrad the applied magnetic field leads to strong deviations from the field-free case. 
For $\phi_f^{(\mathrm{lab})}=0$, for example, a laboratory magnetic field of $B_\mathrm{lab}=2$\,T results in an even stronger focusing of the emitted photons along the ion beam propagation direction compared to the field-free scenario. In contrast, for $\phi_f^{(\mathrm{lab})}=\pi/2$, the magnetic field slightly broadens the angular distribution. 
% \begin{figure*}[t]
%     \centering
%     \includegraphics[width=0.95\textwidth]{Figures/AngularDistribution_Phi0_phi90_plot-alternative.pdf}
%     \caption{The angular distribution of the emitted photons in the $1\text{s}^2$ $^1\text{S}_0$ $ \to$ $1\text{s}2\text{p}$ $^1\text{P}_1 $ scattering process in He-like Calcium for $\gamma=2395$ with and without an external magnetic field, normalized by $\mathrm{d}\sigma/\mathrm{d}\Omega\left(\theta_f=\phi_f=0,B_{Lab}=0\right)=W_0$. The upper plot shows the result in the rest frame of the ion while the lower plot shows the Lorentz transformed result in the laboratory frame. The external electromagnetic fields result in an asymmetry with respect to the azimuthal angle $\phi_f$.}
%     \label{fig: Angular distribution plot}
% \end{figure*}
%
%
% \begin{figure}[t]
%     \centering
%     \includegraphics[width=0.45\textwidth]{Figures/AngularDistribution_Ratio_Phi0_plot.pdf}
%     \caption{The ratio of the angular distribution of the emitted photons for $\gamma=2394$ with and without an external magnetic field. The upper plot shows the result in the rest frame of the ion while the lower plot shows the Lorentz transformed result in the laboratory frame.}
%     \label{fig: Angular distribution plot}
% \end{figure}
%
% \begin{figure}[t]
%     \centering
%     \includegraphics[width=0.45\textwidth]{Figures/AngularDistribution_Ratio_plot.pdf}
%     \caption{The ratio of the differential cross sections of the emitted photons for $B_{lab}\neq 0$ and $B_{lab} = 0$.}
%     \label{fig: Angular distribution ratio plot}
% \end{figure}
%
%
Besides the angular distribution, the polarization of scattered photons can also be strongly affected by the external magnetic field.
Similarly to before, we start the discussion of this $B$-dependence with a short reminder of the field-free case.
Namely, for the here considered scenario of a $J_i=0 \to J_\nu =1 \to J_f=0$ transition and circularly polarized incident radiation, the polarization of scattered light is described by three Stokes parameters~\cite{Volotka22}:
\begin{subequations}
\label{eq: Stokes B=0}
\begin{align}
    P_1^{(f),(\mathrm{ion})}\left(\theta_f^{(\mathrm{ion})},\phi_f^{(\mathrm{ion})}\right)&=\frac{-2\sin^2\theta_f^{(\mathrm{ion})}}{3+\cos 2\theta_f^{(\mathrm{ion})}},\label{Eq.: P1 B=0}\\
    P_2^{(f),(\mathrm{ion})}\left(\theta_f^{(\mathrm{ion})},\phi_f^{(\mathrm{ion})}\right)&=0,\label{Eq.: P2 B=0}\\
     P_3^{(f),(\mathrm{ion})}\left(\theta_f^{(\mathrm{ion})},\phi_f^{(\mathrm{ion})}\right)&=\frac{-4\cos\theta_f^{(\mathrm{ion})}}{3+\cos2\theta_f^{(\mathrm{ion})}}\label{Eq.: P3 B=0}.
\end{align}
\end{subequations}
While these expressions are written in the ion rest frame, their transformation to the laboratory frame follows the simple relation
\begin{equation}
   P_{1,2,3}^{(f),(\mathrm{ion})}\left(\theta_f^{(\mathrm{ion})},\phi_f^{(\mathrm{ion})}\right)=P_{1,2,3}^{(f),(\mathrm{lab})}\left(\theta_f^{(\mathrm{lab})},\phi_f^{(\mathrm{lab})}\right),
\end{equation}
where $\phi_f^{(\mathrm{ion})}=\phi_f^{(\mathrm{lab})}$ and the transformation of $\theta_f$ is given by Eq.~\eqref{eq.: lorentz trafo angle}.\\
Utilizing the theory presented in Sec.~\ref{Section:ResonantScattering} and performing again the averaging over the frequency distribution, we can also calculate the Stokes parameters of light scattered on an ion exposed to external electromagnetic fields.
These Stokes parameters, as observed in the laboratory frame in the uncooled beam scenario, are presented in Fig.~\ref{fig: Polarization Angular Plot} for three different external magnetic field strengths, $B_\mathrm{lab}=0$ (left column), $B_\mathrm{lab}=0.5$\,T (middle column) and $B_\mathrm{lab}=1$\,T (right column).
Moreover, we investigate the cases when the photons are emitted within the y-z-plane ($\phi_f=0$) or perpendicular to it ($\phi_f=\pi/2$), see Fig.~\ref{fig:Field-Geometry}. 
For the field-free case, $B_\mathrm{lab}=0$, the result of our calculations perfectly reproduces the analytical expression~\eqref{eq: Stokes B=0}. 
However, the growth of $B_\mathrm{lab}$ leads to a significant modification of the Stokes parameters.

A very pronounced example of the external magnetic field effect can be observed for the back scattering of photons, $\theta_f^{(\mathrm{ion})}=\theta_f^{(\mathrm{lab})}=0$.
We remind here that the polar angles $\theta_i$ and $\theta_f$, describing the propagation of incident and scattered photons, are defined with respect to the ion beam propagation direction (y-axis). Thus, backscattering corresponds to the case of $\theta_i=\pi$ and $\theta_f=0$, see Fig.~\ref{fig:Field-Geometry}.  
For this case, the field-free formula~\eqref{Eq.: P3 B=0} predicts a complete polarization transfer from initial to final photons, $P_3^{(i)}=1$ to $P_3^{(f)}=-1$.
This transfer of polarization can serve as a promising tool for generating circularly polarized $\gamma$-rays at the Gamma Factory, which are important, for example, for the production of longitudinally polarized muon and positron beams~\cite{Apyan:2022ysh}.
However, as seen from the middle and right panels of Fig.~\ref{fig: Polarization Angular Plot}, the external magnetic field leads to the reduction of $|P_3^{(f)}|$, thus, resulting in partial polarization transfer.
In contrast, the linear polarization of back scattered photons becomes nonzero with increasing $B_\mathrm{lab}$ and can reach quite high values.

\begin{figure}   
    \includegraphics[width=0.45\textwidth]{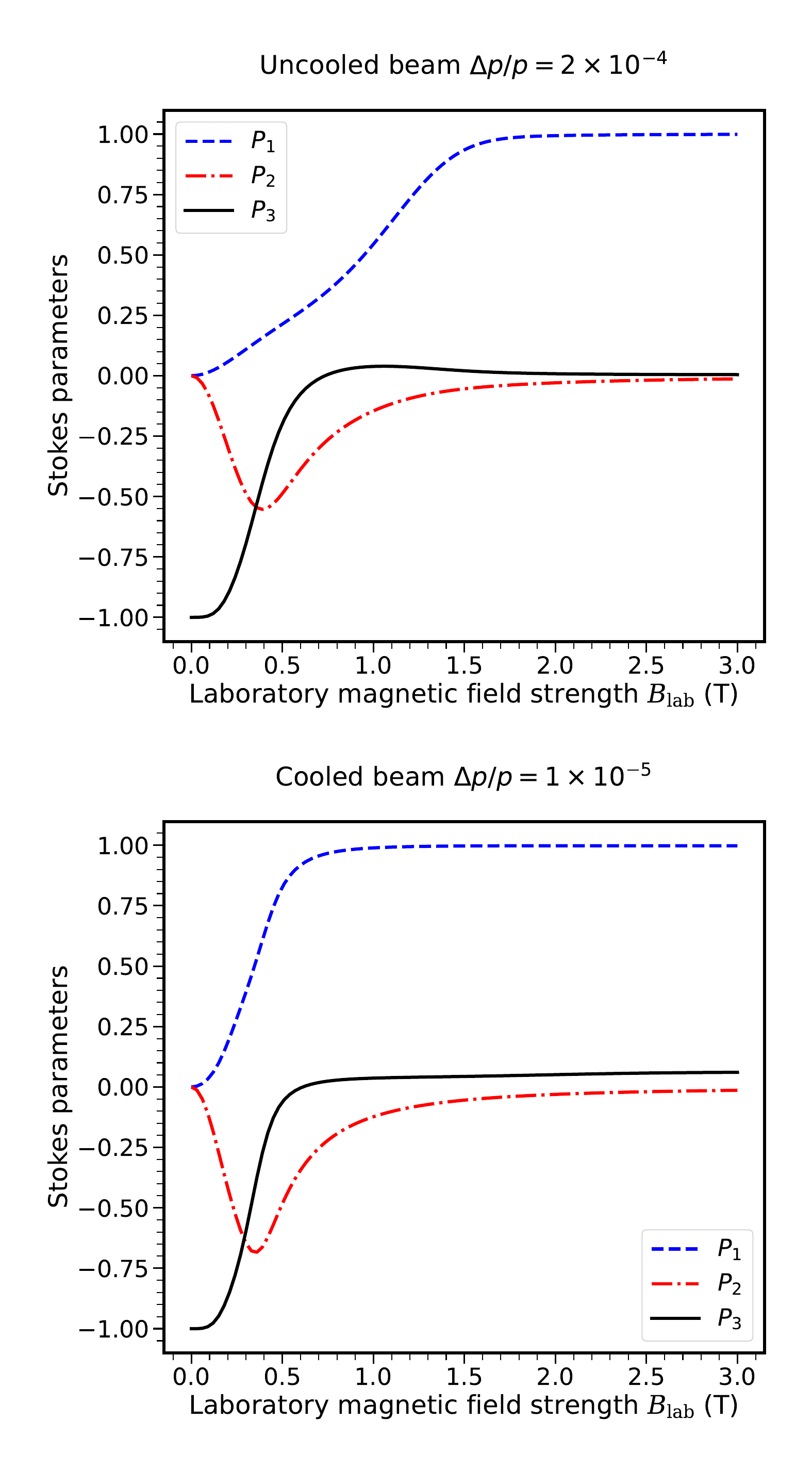}
    \caption{The three Stokes parameters $ P^{(f),(\mathrm{lab})}$ of the emitted photons for $\theta_f^{(\mathrm{lab})}=\phi_f^{(\mathrm{lab})}=0$ as a function of the laboratory magnetic field strength. Calculations were performed for the scenario of an uncooled ion beam (upper plot) as well as for a cooled beam (lower plot).}
    \label{fig: Polarization Plot}
\end{figure}

The $B_\mathrm{lab}$-dependence of the Stokes parameters of back scattered light with $\theta_f^{(\mathrm{ion})}=\theta_f^{(\mathrm{lab})} = 0$ is additionally visualized in Fig.~\ref{fig: Polarization Plot}. 
As seen from this figure, the degree of circular polarization $|P_3^{(f)}|$ rapidly drops with the increase of $B_\mathrm{lab}$, while, in turn, the linear polarization parameters $P_1^{(f)}$ and $P_2^{(f)}$ become non-vanishing. 
In particular the Stokes parameter $P_1^{(f)}$, which describes linear polarization of outgoing radiation within or perpendicular to the scattering plane, almost reaches unity for $B_\mathrm{lab}>0.5$\,T both for the cooled and uncooled beam scenario.
Such a conversion of circular polarization into linear one may offer interesting applications for the Gamma Factory, which will be discussed in the next section.  
%showing a strong depolarization for a laboratory field strength below 1 T. 


\section{Photon Scattering at Gamma Factory}\label{Section:Applications}
%\subsection{Introductory remarks}
While the theory presented in the previous sections is general and can be applied to any storage ring experiment, below we discuss how its predictions can be used to advance the research program of the Gamma Factory (GF) at CERN.
The Gamma Factory scientific program opens new perspectives for 
particle, nuclear, atomic, and applied physics, see Ref.~\cite{Krasny:2023ptc} for further details.
% Less exposed is its capacity, of importance for  the general scientific programme of CERN, 
% (1) to improve the quality of the beams accelerated
% and stored in the CERN circular machines: SPS and LHC,  and (2) to reach  
% an unprecedented precision of their control. 
Less exposed is its capacity to improve the quality of the beams accelerated and stored in the SPS and LHC rings of CERN and to reach an unprecedented precision of their control.
These tasks are of 
%MWK definite 
vital
%MWK
importance for the general scientific program of CERN.

In this section we explore, how the GF project can benefit from investigations of resonant photon scattering off partially stripped ions (PSIs) in the presence of an external magnetic field in the collision zone.
The specific case of a He-like Calcium beam discussed in this paper can be extrapolated to 
other PSI beams. The theoretical framework, 
presented in this paper, is used to assess the precision of the calibration of the SPS/LHC beam momentum as well as of the momentum spread.
In addition, we discuss below the potential role of varying  the 
strength of the magnetic field for the GF beam-cooling 
scheme proposed in Ref.~\cite{Krasny:2020wgx}
and developed further in Ref.~\cite{Kruyt:2024sty}.
%MWK another very recent article which can be cited here:
%@article{Kruyt:2024sty,
%    author = "Kruyt, Peter and Gamba, Davide and Franchetti, Giuliano",
%    title = "{Simulation studies of laser cooling for the Gamma Factory proof-of-principle experiment at the CERN SPS}",
%    doi = "10.18429/JACoW-IPAC2024-MOPS50",
%    journal = "JACoW",
%    volume = "IPAC2024",
%    pages = "MOPS50",
%    year = "2024"
%}
%
Finally, the experimental control of the GF photon beam polarization by an external magnetic field is explored. 

%a paper on Lep momentum calibration
%J.-P. Koutchouk, M. Placidi, C. R. Physique 3 (2002) 1121–1130

\subsection{GF photon-ion collision scheme} 
\subsubsection{Technological challenges}
 The GF beam test demonstrated that PSI beams can be produced, collimated, and stored in the CERN's LHC and SPS rings \cite{Gorzawski:2020dgx,Schaumann:2019evk, Hirlaender:2018rvt}. Their $\gamma_L$, can  
span the values between 15 and 3300. 

In the GF scheme,  PSI bunches collide with photon pulses which can be produced with commercially available, low-phase-noise lasers operating in ultraviolet, visible, or near-infrared regimes.  
These laser pulses are stored in the high-finesse Fabry-Perot cavity to collide them with the PSI bunches at a 20 MHz repetition rate. 
Recent studies demonstrated that the GF laser system can be installed in the high-radiation environment of the CERN accelerator tunnels and confirmed the stable operation of the Fabry-Perot cavity storing light pulses with an average power of 500\,kW~\cite{Mazzola:2024xqq,Lu:2024gwe}. 
These results demonstrate the technological readiness for the implementation of the GF project at CERN.  
The final GF research and development step, the Proof-of-Principle experiment~\cite{Krasny:2019wch}, which will be installed in the SPS tunnel, aims to deliver the final proof of the GF project feasibility, not only in its technological but also in its operational aspects. 
%@article{Krasny:2019wch,
%    author = "Krasny, M. W. and Martens, A. and Dutheil, Y.",
%    collaboration = "Gamma Factory Study Group",
%    title = "{Gamma Factory Proof-of-Principle experiment}",
%    reportNumber = "CERN-SPSC-2019-031, SPSC-I-253",
%    year = "2019"
%}

For collisions of the laser pulses with the PSI bunches in the presence of the magnetic field, as considered in the present study, the Fabry-Perot cavity has to be placed inside the dipole magnet.
Dipole magnets, respecting the requisite aperture and length constraints, and generating the magnetic field of up to 3 Tesla, can be built using the present technology. 

The trajectories of the ions circulating in the SPS/LHC
rings will be perturbed by introducing the magnetic field in the collision zone. The 15 cm long dipole magnet, which generates a 3 Tesla field, introduces an additional 0.45 Tesla-meter
beam bending power. The bending power of the SPS and the LHC dipoles are  12.6 and 120 Tesla-meter, respectively.
The small perturbation of the beam-particle trajectories will have to be compensated by introducing the requisite correction magnets in the vicinity of the collision zone. 

Commercially available instruments measure magnetic fields, of the strength of up to 13 Tesla, with an accuracy better than 10 ppm. The $B$-field dependent effects, discussed in this paper, can thus be measured with high precision. It will be limited only by the calibration accuracy of the PSI beam parameters and by the theoretical calculation precision of the ionic energy levels and the scalar and tensor polarizabilities. 

\subsubsection{Configuration parameters}
 
The specific choice of laser, delivering pulses of photons of a small energy band centered at a chosen value of  $\omega_\mathrm{lab}$, the choice of the atomic transition, characterized by the energy difference between the ground state and excited state of the PSIs, $\omega_\mathrm{ion}$,  and the angle between the Fabry-Perot cavity and the beam axis, $\Theta$, determine the requisite $\beta$ and $ \gamma_L$ of the ion beam which satisfy the resonance condition: 
\begin{equation}
    \omega_\mathrm{ion} = ( 1 + \beta \cos\Theta )\gamma _L \omega_\mathrm{lab}.   
    \label{eq:resonant-energy}
\end{equation}
% The resonant $\gamma_L$ maximizes the rate of produced gamma-ray photons.
The Fourier-limited laser-photon bandwidth  
can be tuned to match, or be smaller than the $\gamma_L$ spread of the PSI beam. The spectral and timing characteristics of the laser pulses and the ion beam optics will be optimized individually for each of the Gamma Factory research domains. 

The length of the dipole magnet has to be larger than the longitudinal size of the PSI bunches which, for the present  SPS and LHC beams, are approximately 15 cm. The PSI bunch length, and the corresponding $\gamma_L$ spread,  can be reduced by the longitudinal cooling of the beam~\cite{Krasny:2020wgx,Kruyt:2024sty}.

The propagation length of the excited Ca-ions in the laboratory reference frame, both at the SPS and LHC energies,  is below 1 cm. It is significantly smaller than the ion bunch length. In such a configuration both the ground state and excited level energy shifts have to be taken into account while calculating the energy flow of the emitted $\gamma$-rays. 

The $\Theta$ angle for all the considered GF applications ranges between 1 and 2.6 degrees. The maximal $\Theta$ angle and the PSI bunch length determine the minimal aperture of the dipole magnet.
One should note, that in our geometry, displayed in Fig.~\ref{fig:Field-Geometry}, the polar angle of the incident radiation is $\theta_i = \pi - \Theta = \pi$, while in reality the angle will be slightly smaller, $\theta_i<\pi$. However, as discussed in Ref.~\cite{Serbo22}, the effect of a small but nonzero angle $\Theta$ on the scattering process is negligible and, hence, is not considered in our calculations presented in the previous sections.  
%MWK
%Jan, I am not sure if I understand what you want to say above:
% The actual value of \Theta has to be taken into account if you want to calibrate the absolute 
%momentum scale of the PSI bem to the precision of 10^-5 .  It is true that for the typical GF values 
% it will not modify, polarisation, angular distributions in the visible way...but has to 
%be taken into account as depicted in formulae (\label{eq:resonant-energy}.
%MWK

\subsection{Gamma Factory Applications}
\subsubsection{Resonance condition tuning} 

The energy differences between the excited  $1\text{s}2\text{p}$ $^1\text{P}_1 $  and the ground state $1\text{s}^2$ $^1\text{S}_0$ of He-like PSIs, can be calculated 
with a precision better than 0.01~\%~\cite{Yerokhin2019}. For the transition in He-like Ca ions, considered in this paper, the calculated~\cite{Yerokhin2019} and the measured \cite{Seely1989,rice2014} values agree within the present measurement errors which are also below 0.01~\%. 
The $\gamma_L$ of the SPS beams   
is presently calibrated 
with the precision of 0.02 \% \cite{Wenninger:2005ir}.
However, the absolute beam energy uncertainty increases 
to 0.1~\% \cite{Todesco:2017nnk} at the top LHC beam energy and becomes significantly
larger than the expected theoretical precision of the transition frequency.   
As a consequence, $\gamma_L$ of the beam has to be tuned in a dedicated beam-energy scan to find its optimal value satisfying the resonance condition~(\ref{eq:resonant-energy}). For the canonical high-energy LHC beams, having the $\gamma_L$ bandwidth of $2 \times 10^{-4}$, the scan has to be made in approximately 40 beam momentum setting steps covering $\pm 2\sigma$ beam momentum uncertainty interval, with the step size of $10^{-4}$ of its nominal value. Such a scan requires changing, in each step,  the PSI beam optics over the whole ring circumference, which will be both time-consuming and cumbersome. 

The resonance tuning procedure can be vastly simplified by replacing the $\gamma_L$ scans with the magnetic field scan at a fixed 
$\gamma_L$ value.
For the LHC He-like Ca beam, the Stark shift of the 
excited energy level for $B_\mathrm{lab}=3$\,T is of the order of  0.2 \% of its nominal value. It is thus 2 times larger than the initial beam-energy ($\gamma_L$) calibration uncertainty. By choosing the initial setting of the $\gamma_L$ as the calculated mean value for its $B_\mathrm{lab}=0$ and $B_\mathrm{lab}=3$\,T predictions,  and by step-wise decreasing $B_\mathrm{lab}$ in the collision zone from the value of 3 Tesla down to zero,  the peak in the production rate of the GF photons can be found and used for the fast and simple determination of the resonant value $\gamma_L$. 

\subsubsection{LHC beam-energy calibration}

\begin{figure*}
    \centering
    \includegraphics[width=\textwidth]{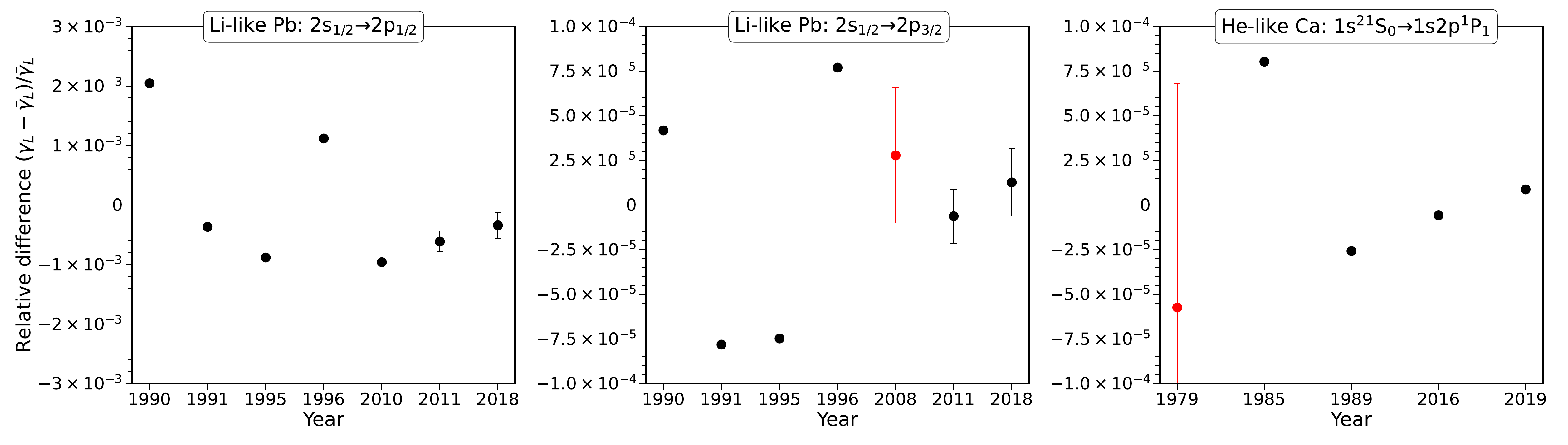}
    \caption{
    The relative difference between the resonant value of the Lorentz factor $\gamma_L$, derived from calculated (black) and measured (red) transition energies, to their average value $\bar\gamma_L$. The transitions considered are $2\mathrm{s}_{1/2}\to 2\mathrm{p}_{1/2}$ and $2\mathrm{s}_{1/2}\to 2\mathrm{p}_{3/2}$ in Li-like Pb~\cite{Indelicato1990,kim1991resonance,chen1995relativistic,johnson1996transition,kozhedub2010relativistic,sapirstein2011s,yerokhin2018energy,Zhang2008EBIT}, and $1\text{s}^2$ $^1\text{S}_0$ $ \to 1\text{s}2\text{p}$ $^1\text{P}_1 $ in He-like Ca~\cite{Sugar1979,Vainshtein_1985,Seely1989,Si2016,Yerokhin2019}. The Lorentz factors are calculated for a fixed laser photon energy of 0.815 eV and an angle $\Theta = 1^\circ$.}
    \label{fig:Gamma Calibration plot}
\end{figure*}

Once the resonance finding step is finalized, the precision of the $\gamma_L$ calibration reflects the present accuracy of calculations of the involved energy levels. As illustrated in Fig.~\ref{fig:Gamma Calibration plot} for Li-like Pb and He-like Ca, this results in an uncertainty in the order of $10^{-4}$. This corresponds to a tenfold improvement in the beam energy calibration precision of the LHC beams, becoming the most precise method of energy calibration of the high-energy hadronic beams.  

%MWK
%MWK The plot I suggest here would be, perhaps the most important plot of this paper for the cacclerator community. It should show on the y axis the resonant 
%MWK \gamma_L resonant value for a fixed \omega and  \Theta
%MWK as a function (x-axis) of the earlier measurements  and calculation results  determining the requisite  \omega_i values for the calibration procedure. 

Building on the results from the previous sections, we propose a complementary calibration method. For ions with scalar and tensor polarizabilities known with sufficient accuracy, the $\gamma_L^2$ dependence of the Stark shift can be exploited. The high-precision calibration procedure involves relative $B$-field-dependent adjustments of the laser photon energy $\omega^\mathrm{res}_\mathrm{lab}(B_\mathrm{lab})$, maintaining the resonance condition with the $\ket{^1\mathrm{P}_1,M_\nu=\pm1}$ sublevel at each $B$-field strength setting. Using Eq.~\eqref{Eq.: Zeeman and Stark shift 1S0 and 1P1}, we derive the expression
\begin{align}
  &\gamma_{L,calib} \label{eq:calibrated_gamma_L}\\
  &=   \frac{( 1 + \cos\Theta ) \left(\omega ^{res}_{lab} (B_{\mathrm{lab}}) - \omega ^{res}_{lab} (B_{\mathrm{lab}}=0)\right)}{(\frac{1}{4} \alpha_2\left(^1\text{P}_1\right) - \frac{1}{2} \alpha_0\left(^1\text{P}_1\right) + \frac{1}{2} \alpha_0\left(^1\text{S}_0\right) )  c ^2 B_{\mathrm{lab}}^2}, \nonumber   
\end{align} 
where for high $\gamma_L$ we set $v\approx c$ and hence $\beta \approx 1$. 
The calibrated $\gamma_{L,calib}$ is obtained as the average over all B-field strength settings.
To eliminate the contribution of the scalar polarizabilities, one can take, for each B-field strength, the difference of the resonant laser photon energy for the $\ket{^1\mathrm{P}_1,M_\nu=0}$ and the $\ket{^1\mathrm{P}_1,M_\nu=\pm1}$ sublevels. 
In this case, only the tensor polarizability contributes and the expression simplifies to:
\begin{align}
&\gamma_{L,calib} \label{eq:calibrated_gamma_L}\\
  &=   \frac{( 1 +  \cos\Theta ) \left(\omega ^{res}_{lab} (M_\nu=\pm1) - \omega ^{res}_{lab} (M_\nu=0)\right)}{\frac{3}{4} \alpha_2\left(^1\text{P}_1\right)   c ^2 B_{\mathrm{lab}}^2}. \nonumber   
\end{align}

The precision of this method is no longer constrained by the accuracy of the energy level calculations but instead by the precision of the polarizability calculations. If calculations of transition energies or polarizabilities are sufficiently accurate, the calibration precision could be improved to $10^{-5}$, achieving the same level of precision as the LEP electron/positron beams.    
\subsubsection{Measurement of polarizabilities of highly charged ions}
The calibration procedure described earlier relies on the precise knowledge of the scalar and tensor polarizabilities of the involved states. However, as proposed in Ref.~\cite{Budker2020}, if these polarizabilities are not known with sufficient accuracy, the method can be turned around to extract these values instead. The strong enhancement of the electric field in the ion frame and the resulting visible Stark shift opens the possibility to study polarizabilities for medium-Z highly charged ions.

\subsubsection{Beam cooling and its experimental control }
\label{Sec: Beam cooling}
\paragraph{Cooling scheme\\}
Laser beam cooling~\cite{Eidam:2017csp, Krasny:2020wgx,Kruyt:2024sty} is based on the selective excitation of only a fraction of ions stored in the PSI bunches.
By tuning the spectral width of laser-photon pulses such that it is smaller than the PSI ion momentum smearing and detuning the resonant $\gamma _L$ to lower values, only the fastest ions of the ion-bunch can resonantly absorb photons. 
This cooling process involves a gradual increase of $\gamma_L$ until also the slowest ions are excited. 
However, introducing a magnetic field in the ion-photon collision zone allows the replacement of the technically challenging $\gamma _L$ ramp by a magnetic field strength ramp. In this scenario, $\gamma_L$ can be chosen such that, for $B_{lab}=3$\,T only the fastest ions are excited to the $\ket{^1\mathrm{P}_1,M_\nu=\pm1}$ sublevel. By decreasing the magnetic field strength, the resonant transition frequency of the $\ket{^1\mathrm{P}_1,M_\nu=\pm1}$ sublevel can be lowered to excite slower ions as well.
One may note, however, that the magnetic field dependent cooling is by a factor of $\sim2$ slower than the $\gamma _L$-ramp cooling, as only the $\ket{^1\mathrm{P}_1,M_\nu=\pm1}$ sub levels contribute to the cooling process (see Fig.~\ref{fig:Total Cross section over omega plot}).

\begin{figure}
    \centering
    \includegraphics[width=0.95\linewidth]{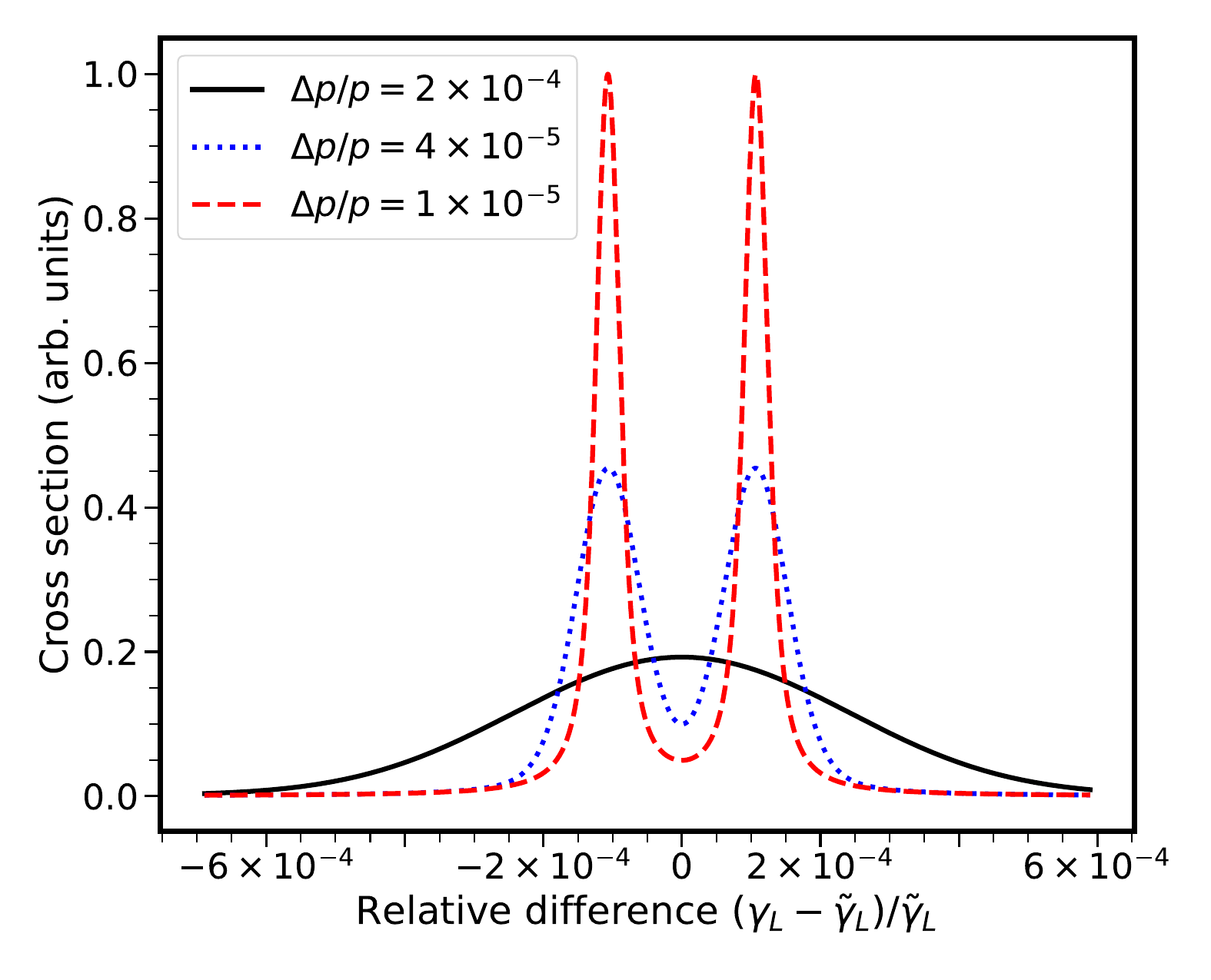}
    \caption{The cross section of photons scattered in a finite solid angle in y-direction with an opening angle of 1\,mrad in the laboratory frame as a function of the relative difference of the Lorentz factor $\gamma_L$ with respect to $\tilde\gamma_L$, corresponding to the center of the $M_\nu=\pm1$ sublevel. Calcualtions were performed for different momentum spreads of the ion beam, showing how the Zeeman splitting of the $M_\nu=\pm1$ sublevel becomes visible for a sufficiently cooled beam.}
    \label{fig:Zeeman_spliting of gamma}
\end{figure}

\paragraph{Control of the momentum spread\\}
The achieved degree of longitudinal beam cooling,
observed as the reduction of the energy-spread of the PSI beams, can be 
monitored by observing the appearance of the Zeeman spliting of the resonant $\gamma _L$ values 
as shown in Fig.~\ref{fig:Zeeman_spliting of gamma}.
%MWK The plot suggested here is a variant of the Fig.4b plot for B=3 Tesla.
%WMK on the x-axis $\gamma _L$ would replace $\omega _ i $.
%MWK and only a fraction of the x axis would be shown (corresponding
%MWK o the values of \omega _i between 3909 and 3912 eV
In this plot the Zeeman splitting is shown for three values of the PSI beam momentum spread: for the nominal non-cooled one, and
for the beam with reduced Gaussian width of the 
beam energy spread by a factor of 5 and 20.
The characteristic double-peak structure  invisible for  the 
initial momentum spread of the LHC beams becomes visible
as the cooling of the beam progresses. The measured shape of the 
resonant $\gamma _L$ distribution for $B=3$\,T will determine 
the spectral distribution of the PSI beam at consecutive stages 
of the beam cooling procedure.

\subsubsection{GF photon beam polarization}

Several domains of the GF experimental program require 
the precise polarization control of the generated photon beams. 
Circular polarization, for instance, is crucial for the production of longitudinally polarized leptons~\cite{Apyan:2022ysh} or potential atomic parity violation studies~\cite{Richter2022}. 
Meanwhile, linearly polarized $\gamma$-rays can be used for studies of vacuum birefringence~\cite{Budker:2021fts,Karbstein:2021otv} and Delbrück scattering~\cite{Budker2020,Sommerfeldt2023}.
% Linearly polarized $\gamma$-rays lead to azimuthal asymmetries of produced leptons which can be precisely measured and used to determine the degree of linear polarization of the photons~\cite{Gros:2016dmp}.
%However, the degree of circular polarization of the photons cannot be deciphered from the angular distribution of such produced leptons.
As discussed in Sec.~\ref{sec: angular distribution and polarization}, circularly polarized $\gamma$-rays can be produced in collisions of circularly polarized laser photons with He-like ions due to perfect polarization transfer.
We demonstrated that the external laboratory magnetic field has a strong impact on this transfer, see Figs.~\ref{fig: Polarization Angular Plot} and \ref{fig: Polarization Plot}. For instance, the applied $B_\mathrm{lab}$-field can transform circular polarization into linear one, offering additional control over the polarization of scattered radiation. This opens up the possibility of tailoring the polarization of $\gamma$-rays to meet specific experimental requirements on demand.
% As shown in Fig.\ref{fig: Polarization Plot}, the external magnetic field transforms circular polarization into linear polarization, providing a unique opportunity to convert the measurement of the degree of circular polarization into the measurement of linear polarization.
% Indeed, Fig.~\ref{fig: Stokes 2 Plot} shows the Stokes parameter $P_2^{(f)}$ of the emitted photons for $\theta_f=\phi_f=0$ as a function of $B_\mathrm{lab}$, revealing a strong sensitivity to the initial degree of circular polarization.
% Hence, the degree of circular polarization for the $B=0$ case can 
% be inferred from the measured B-dependent evolution of the Stokes parameter $P_2$ of the $\gamma$-rays produced in the 
% B-field collision zone. 
% \begin{figure}   
%     \includegraphics[width=0.45\textwidth]{Figures/Stokes2Plot.pdf}
%     \caption{The second Stokes parameter $P_2^{(f)}$ of the emitted photons for $\theta_f=\phi_f=0$ as a function of the laboratory magnetic field strength. Calculations were performed for different degrees of circular polarization of the incoming radiation.}
%     \label{fig: Stokes 2 Plot}
% \end{figure}

%showing a strong depolarization for a laboratory field strength below 1 T. 
%
%
%
%
%
%
\section{Conclusion}\label{Conclusion}
%Witek's suggestions
We presented a theoretical study on resonant photon scattering by partially stripped ions exposed to external electric and magnetic fields.
Special emphasis was placed on the geometry in which laser photons collide with counter-propagating relativistic ion beams.
We assume moreover that a dipole magnet is placed in the collision zone, leading to strong electric and magnetic fields in the ion rest frame as a result of the Lorentz transformation.
For this experimental setup, we investigated the effect of external fields both on the scattering rate and on the angular distribution and polarization of outgoing photons.\\
The developed theoretical approach was applied to explore in detail the resonant photon scattering by He-like Ca$^{18+}$ ions for magnetic field strengths and collision energies typical for the GF project at CERN.
Detailed calculations indicate a strong impact of the laboratory magnetic field on the rate of detected scattered photons as well as on their polarization which arises due to Zeeman and Stark shifts of the ionic states. In particular, we showed that the rate of detected photons decreases with the growth of $B_\mathrm{lab}$, depending on the energy spread off the ion beam. 
Moreover, we demonstrated that an external magnetic field can be utilized to convert initially circularly polarized photons into linearly polarized ones. \\
Based on the result of our calculations, we argue that the magnetic field sensitivity of the resonant scattering process can be of importance for various aspects of the GF project.
For example, the $B_\mathrm{lab}$-dependence of the detection rate and the visible Stark splitting of the excited sublevels can be used to facilitate the resonance condition tuning and offers a new beam cooling scheme.
Furthermore, the laboratory magnetic field provides precise and flexible polarization control of the outgoing photon beam.

\acknowledgments

J.R. and A.S. acknowledge funding by the Deutsche Forschungsgemeinschaft (DFG, German Research Foundation) under Germany’s Excellence Strategy – EXC-2123 QuantumFrontiers – 390837967.

\FloatBarrier
\bibliographystyle{ieeetr}
\bibliography{main.bib}
%\printbibliography
\end{document}